\documentclass[preprint,11pt]{elsarticle}
\usepackage[leqno]{amsmath}
\usepackage{amssymb}
\usepackage{amsthm}
\usepackage{amsfonts}
\usepackage{color}
\usepackage{graphicx}
\usepackage{booktabs}
\usepackage{graphicx,epstopdf}
\usepackage{mathtools}
\usepackage{mathrsfs}
\usepackage{multirow}
\usepackage{rotating}
\usepackage{enumerate}
\usepackage{epsfig}
\usepackage{float}
\usepackage{setspace}
\usepackage{ragged2e}
\usepackage{blindtext}
\usepackage{lineno}
\usepackage{url}
\usepackage{xcolor}
\usepackage{url}
\usepackage[colorlinks=true, linkcolor=black, citecolor=black, urlcolor=black]{hyperref}
\usepackage{float}
\usepackage{lipsum} 
\usepackage{verbatim}

\newtheorem{thm}{Theorem}
\newtheorem{rem}{Remark}

\usepackage{etoolbox}
\usepackage{caption}
\usepackage{subcaption}
\newtheorem{lemma}{Lemma}

\theoremstyle{definition}

\usepackage{etoolbox}
\usepackage{geometry}
\numberwithin{equation}{section}
\AtBeginEnvironment{thebibliography}{\small}
\begin{document}

\begin{frontmatter}

\title{An additional food driven biological control patch model, incorporating generalized competition}

\author{Urvashi Verma$^{1}$}
\ead{uverma@iastate.edu}
\author{Kanishka Goyal$^{1}$}
\ead{kgoyal@iastate.edu}
\author{Chanaka Kottegoda$^{2}$}
\ead{kottegoda@marshall.edu}
\author{Rana D. Parshad$^{1}$\corref{my corresponding author} }
\ead{rparshad@iastate.edu}
\address{1) Department of Mathematics, Iowa State University, Ames, IA 50011, USA \\
2)Department of Mathematics and Physics, Marshall University,
Huntington, WV 25755, USA }

\cortext[my corresponding author]{Corresponding author}



\begin{abstract}
Additional food sources for an introduced predator are known to increase its efficiency on a target pest. In this context, inhibiting factors such as interference, predator competition, and the introduction of temporally dependent quantity and quality of additional food are all known to enable pest extinction. As climate change and habitat degradation have increasing effects in enhancing patchiness in ecological systems, the effect of additional food in patch models has also been recently considered. However, the question of complete pest extinction in such patchy systems remains open. In the current manuscript, we consider a biological control model 
where additional food drives competition among predators in one patch, and they subsequently disperse to a neighboring patch via drift or dispersal. We show that complete pest extinction in both patches is possible. Further, this state is proved to be globally asymptotically stable under certain parametric restrictions. We also prove a codimension-2 Bogdanov-Takens bifurcation. We discuss our results in the context of designing pest management strategies under enhanced climate change and habitat fragmentation. Such strategies are particularly relevant to control invasive pests such as the Soybean aphid (\emph{Aphis glycines}), in the North Central United States.
\end{abstract}

\end{frontmatter}

\section{Introduction}
Invasive species and pests are a major global concern, causing substantial annual crop damage \citep{paini2016global,pimentel2005update}, and posing serious environmental and economic threats \citep{TS14, CO18, BOP22}. Managing them is challenging, and while there are several control strategies, chemical treatments constitute a significant share of these \citep{PB14}. This is despite their negative environmental effects, such as pest species developing resistance \citep{hladik2018environmental, bass2015global}. Thus, there is a need for alternative strategies that are eco-friendly, such as top-down or classical biological control. This method involves introducing a natural enemy of the targeted pest species \citep{V96, S99}, to suppress the pest population, thereby decreasing the frequent use of insecticides \citep{heimpel2013environmental}. In cases when the introduced predator fails to reduce the pest population to the desired level, its effectiveness can be improved through several methods, one of which is providing additional food (AF) to the predator, different from the targeted prey \citep{SV06, T15}. Several mathematical models have been formulated to describe predator-pest dynamics with AF \citep{SP07, SP10, S02, SP11, SPV18, SPD17, VA22}, suggesting that providing a sufficient quantity of high-quality AF to the predator will result in pest extinction. However,  AF-mediated models with species movement mechanisms such as dispersal and drift have been far less investigated.

Climate change can have direct and indirect influences on species movement \citep{travis2013dispersal}, since the decision to disperse can be affected by changes in the speed of wind \citep{thomas2003aerial}, in storm conditions \citep{lea2009extreme}, and in flooding events \citep{roche2012flooding}. This type of movement mechanism can be categorized as drift, where the flow of water/direction of wind determines the direction of movement, typically from an upstream to a downstream location. Species can also disperse for food, mates, and resources \citep{CL03}, and several studies incorporate dispersal in prey-predator systems, including dispersal in only one species and dispersal linked to predation \citep{arditi2018asymmetric,messan2015two}. Fragmentation of natural habitat due to human intervention will result in a growing number of habitat ``patches", and species will often disperse between these small interactive ``patches." To this end, patch models have been intensely investigated \citep{W95, T11} as the dynamics of prey-predator systems in a two or multi-patch setting with dispersal between patches can differ from those where only a single patch is considered \citep{GK05, BSC10}. Recent research has also explored drift in aquatic ecosystems with network structures, so in multi-patch environments, as well as in eco-epidemic systems with dispersal \citep{chen2024evolution, chen2023evolution, salako2024degenerate}. 
Evidence from the landscape ecology literature suggests that natural enemies of pests are more abundant in smaller patches adjacent to crop fields \citep{HL20}. Spatial arrangements such as landscape heterogeneity and providing alternative food sources to natural enemies are promising approaches not only for biological pest control, but they also support biodiversity conservation \citep{landis2000habitat, KM18, klinnert2024landscape}. This combined strategy is often referred to as ``landscape features supporting natural pest control" (LF-NPC). Among other landscape management practices, one is the strategy of planting prairie strips, small
sections (about 10-20 $\%$) of crop fields dedicated to various plants instead of the primary crop. STRIPS (Science-based
Trails of Row Crops Integrated with Prairie Strips) is an ongoing project pioneered and led in Iowa \citep{S16} that explores how
integrating a small percentage of prairie into row crops significantly improves soil and water quality, boosts biodiversity,
and is among the most affordable conservation practices for farm landowners \citep{LS17}.
Literature suggests that this AF source `` boosts'' predators' energy, driving their dispersal out of the AF patch, into the crop field \citep{carter1984honeydew,monsrud1999aggregative}. Conversely, the AF can also act as an attractant for predators wanting to disperse out of the crop field into the AF patch, particularly if the AF provides better nutritional diversity \citep{borg1999value}. Thus, the inclusion of dispersal and drift in prey-predator patch models with an AF source is increasingly relevant, as it strongly affects species distribution,
population dynamics, and overall ecosystem functioning \citep{M03, D99} - particularly under enhanced effects of climate change and habitat fragmentation. Biological control strategies are effective, but they have limitations. These include the risk of an unbounded predator population \citep{PWB20} and effects on non-target species \citep{PQB16}. In these situations, competition between predators can act as a natural self-regulating system \citep{klomp1964intraspecific}. Prairie strips adjacent to crop fields offer predators nutritious resources like nectar and pollen, along with suitable microclimatic conditions and refuge \citep{kordbacheh2020strips,schulte2017prairie, snyder2019give}.  However, these strips are usually much smaller than the crop fields and exist in limited patches. This can lead to predators gathering in the strips, increasing local predator density. As a result, competition for resources in the prairie or space limitations may increase intraspecific competition among predators. A recent study explored intraspecific competition in the presence of additional food under type IV response \citep{prakash2025role}. 

The following \emph{general} model for an introduced predator population $y(t)$ preying on a target pest population $x(t)$, also provided with an additional food source of quantity $\xi$ and quality $\frac{1}{\alpha}$, has been proposed in the literature \citep{SP07,SP10,SP11}, 
\begin{equation}
\label{Eqn:1g}
\frac{dx}{dt} = x\left(1-\frac{x}{\gamma}\right) - f(x,\xi,\alpha) y, \    \frac{dy}{dt} =  g(x, \xi, \alpha) y  - \delta y.
\end{equation}
Here, $f(x,\xi,\alpha)$ is the functional response of the predator, which depends on both pest and additional food. Likewise,
$g(x,\xi,\alpha)$ is the numerical response of the predator. Earlier literature on AF has shown that increasing $\xi$ beyond a threshold $\xi_{critical}$ in models of type \eqref{Eqn:1g} yields pest extinction. Due to the positivity of solutions, trajectories will converge onto the predator axis, and this can occur in minimal time \citep{SP10, SP11, VA22,prakash2023stochastic}. Subsequently, it was shown that pest extinction only occurs in infinite time \citep{PWB20}, and can also result in infinite time blow-up of the predator population \citep{PWB23}.
The type III response has been considered in a number of AF models, \citep{SPV18, SPD17}, but an adaptation of the results in \citep{PWB20} shows pest extinction can occur asymptotically but only at the cost of unbounded growth of the introduced
predator. 
The model defined in equation \ref{Eqn:1g}, with type II response and constant predator harvesting with rate $\rho$, is studied in \citep{sen2015global}. If $\rho>0$, then up to two interior equilibria can exist, with an unstable axial pest-free state, and finite time extinction of the predator is achievable for certain initial conditions. Several codimension-one bifurcations exist (Hopf and saddle-node) along with a codimension-two Bogdanov-Takens bifurcation. However, the behavior for large initial data above the stable manifold of the interior saddle equilibrium remains unresolved. 
To prevent unbounded predator growth, several predator-dependent inhibitory mechanisms have been proposed. These include prey defense through type IV functional response \citep{VA22}, predator interference by the Beddington-DeAngelis functional response \citep{S02, W23}, purely ratio-dependent response \citep{S45, S46}, and intraspecific predator competition \citep{PWB23}. In Prasad et al. \citep{S02}, the model was considered with the Beddington–DeAngelis functional response \citep{S33} that incorporates mutual interference. In case of high interference, there is always one unique interior equilibrium - if a feasible pest-free equilibrium exists, it is a saddle, making pest eradication unfeasible. For low interference, predator density remains bounded, and up to one interior equilibrium exists depending upon the quantity of AF. Pest-free and interior equilibria may coexist, with the pest-free state being always a saddle. A predator-free equilibrium can also exist, and a pest-free state is globally stable when the quantity of additional food exceeds a certain threshold. Recent work \citep{W23} shows pest extinction is achievable in a tighter parametric regime.

For many decades, people have used bifurcation theory as a tool in understanding how variations in a system's parameters can cause qualitative changes in the dynamical system \citep{kuznetsov1998elements, perko2013differential, guckenheimer2013nonlinear}. Many researchers, \citep{zhu2003bifurcation, xiao2001global}, have studied and analyzed complex predator-prey models with nonlinear functional responses, and have proven that these models often exhibit rich bifurcation structures, including Hopf, saddle-node, and Bogdanov–Takens (BT) bifurcations. In particular, BT point acts as an organizing center from where critical bifurcations emerge, such as homoclinic loops and multiple limit cycles \citep{dumortier2006bifurcations, lu2021global, huang2014bifurcations}. Even a small perturbation in parameters can flip the system's behavior from stable co-existence to oscillations or even to extinction states. In complex predator–prey systems, these rich bifurcation structures help to predict changes in regime and design biologically feasible pest management strategies. Of particular interest are bifurcations of \emph{higher} codimension, wherein qualitative behavior of a system changes as two or more parameters vary. For instance, \citep{banerjee2024bifurcations}, in a generalist predator model with alternative food source, identify degenerate Hopf (codim-3) and Bogdanov–Takens (codim-4) bifurcations; \citep{huang2016bogdanov} establish a codim-3 BT bifurcation in a constant-yield harvesting setting; \citep{zhou2024bogdanov} find Hopf (codim-2) and BT (codim-2 and 3) bifurcations in models with type II response and predator release; and \citep{arsie2022predator, arsie2023high} uncover nilpotent cusp (codim-3), degenerate Hopf, and heteroclinic (codim-2) bifurcations in predator–prey systems with Allee effects under generalized type III and IV functional responses. However, in the context of AF models, the literature has results on primarily codimension one bifurcations \citep{SP07, SP10, SP11} - that is, qualitative changes as the quantity or quality of the AF vary, or perhaps predator birth or death rates, or prey carrying capacity vary.

Competition, among predators, is ubiquitous in natural systems \cite{chase2002interaction}. Previous studies have shown that including predator competition in predator-prey models can generate much richer dynamics than without such competition \citep{bazykin1998nonlinear,huang2014bifurcations,lu2021global}. In \citep{verma2023t}, a two-patch prairie-crop field model with predator movement was considered. Sufficient drift and dispersal were found to prevent predator blow-up, and with drift, the pest extinction in the crop field was proved to be globally stable under certain parametric conditions. Overall pest densities in the two-patch system were lower than in classical AF-only or classical predator–prey models; however, the complete pest extinction state in both patches could not be achieved with either drift or dispersal. Thus, although such movement is crucial to many ecological processes, and heightened due to climate change and habitat fragmentation, complete pest extinction has not been achieved in AF (multi) patch models thus far - it is only seen in single patch models, primarily with interference mechanisms, and competition. Also, higher codimension bifurcations for AF models are much less reported in the literature, to the best of our knowledge \cite{PWB23}.

Motivated by these findings, we consider a two-patch model where intraspecific competition among predators initiates because of the presence of additional food in the AF patch (prairie strip), and the other patch (crop field) follows the classical prey-predator dynamics. Note, although classical predator competition is modeled as a $-y^{2}$ term, in the predator state ($y$), we consider $-y^{p}$, $1<p\leq 2$. This enables us to generalize the competition term to cover applications such as hyperbolic mortality, nonlinear harvesting, generalized competition/interference \cite{sambath2016stability, fenberg2008ecological, antwi2020dynamics, barman2023two}. The special case $p=2$ covers the case of classical competition.
The mathematical analysis for the drift model is provided in Section \ref{drfit_model}, and the analysis for the dispersal model is detailed in Section \ref{dispersal_section}. Section \ref{bifurcation_Section} presents bifurcation analysis of \emph{higher} codimension. Section \ref{Discussion and Conclusion} discusses future directions and summarizes the main findings with biological implications.

Our primary contributions in the current manuscript are as follows,
\begin{itemize}
\item Two novel AF biological control models with patch structure \eqref{eq:model_drift_cann} and \eqref{eq:patch_model2}, are introduced. Here, the additional food triggers generalized intraspecific competition among predators in the AF patch (prairie strip). The predators then move to the neighboring crop field patch via drift, \eqref{eq:model_drift_cann} or dispersal \eqref{eq:patch_model2}. 
\item Complete pest extinction state in both patches is proved to be globally asymptotically stable with both drift and dispersal (see Theorem \ref{thm:E1_global_unidirectional} and Theorem \ref{e1_patch1_dispersal}), respectively.
\item  Pest extinction in the crop field is globally asymptotically stable with drift via Theorem \ref{thm:E2_global_unidirectional}.
\item The patch model enables pest extinction in the crop field with dispersal via Lemma \ref{lem:E2_stability_dispersal}.

\item We also consider a single-patch system \eqref{model}, where we show the existence of the codimension-2 Bogdanov-Takens bifurcation (see Theorem \ref{thm:BT2}). This bifurcation signifies a critical threshold for qualitative changes in predator-prey dynamics, including the emergence of oscillatory behavior in both pest and predator populations.

\item We discuss the implications of our results in controlling certain key invasive pests, such as the soybean aphid (\emph{Aphis glycines}), in the North-Central United States.
\end{itemize}
\section{Modeling Drift Between \texorpdfstring{$\Omega_1$ \& $\Omega_2$}{Omega1 and Omega2}}
\label{drfit_model}
We next derive the biological control model, considered in the current manuscript.
Motivated by pest management tactics and strategies, such as the STRIPS program, we consider two patches that make up our landscape: a crop field ($\Omega_2$), the larger unit, and a prairie strip ($\Omega_1$), the smaller unit. The prairie strip, by design, possesses row crops that provide additional food, such as nectar and pollen, to the predator, which would enhance its effectiveness in targeting the pest that resides primarily in the crop field. The introduction of additional food initiates the competition among predators with an intraspecific rate $c$. The function $f(\xi) = \xi$ drives the competitive interactions amongst predators. The crop field has no AF. Because of the different in sizes of patches $\Omega_1$ and  $\Omega_2$, we have two different carrying capacities $k_p$ and $k_c$. Thus, for the patch $\Omega_1$, the quantity of AF is $\xi$ and the quality of AF is $\frac{1}{\alpha}$. So, in $\Omega_{1}$, we have an AF-driven predator-pest system with pest density as  $x_{1}$ and predator density as $y_{1}$. In $\Omega_2$, we have the classical prey-predator system since $\xi=0$, and the pest density is  $x_{2}$ and predator density is taken as $y_{2}$. The functional response chosen here is the classical Holling type II functional response. 
\begin{equation} \label{eq:model_drift_cann}
\begin{aligned}
 \Dot{x_1}& = x_1 \left (1-\frac{x_1}{k_p}\right )-\frac{x_1 y_1}{1+x_1+\alpha \xi} \\
 \Dot{y_1}& = \epsilon_1 \left ( \frac{x_1+\xi}{1+x_1+\alpha \xi} \right)\ y_1  - q_1 y_1 -c \xi y_1^p, \   \quad 1< p \leq 2\\
 \Dot{x_2}& = x_2 \left (1- \frac{x_2}{k_c} \right)-\frac{x_2  y_2}{1+x_2} \\
\Dot{y_2}& = \epsilon_2 \left ( \frac{x_2}{1+x_2} \right ) \ y_2 - \delta_2 y_2 + q_2 y_1 
\end{aligned}
\end{equation}
The assumption on drift rates is that the predators will drift away from $\Omega_{1}$ into $\Omega_{2}$. The rate of drift out of the prairie strip ($\Omega_{1}$) is $q_{1}$, and the rate at which predators ``arrive" or are drifted into crop field ($\Omega_{2}$) is $q_{2}$. Here, the drift is driven by flooding or wind. 

\subsection{Mathematical Analysis}  
\begin{thm}
Assume the parameters $k_p,k_c,\epsilon_1,\epsilon_2,\xi,\alpha$ are all positive and the drift rates $q_1,q_2$ are non-negative. Then, the model \eqref{eq:model_drift_cann} is positively invariant in $\mathbb{R}_+^4$.
\label{pos_inv_drift}
\end{thm}
\begin{proof}
See Appendix \eqref{proof:pos_inv_drift}.
\end{proof}
We now consider the existence and local stability analysis of the biologically relevant equilibrium points for the system \eqref{eq:model_drift_cann}. The Jacobian matrix $(\hat{J})$ for \eqref{eq:model_drift_cann} is given by: 
\begin{equation} 
 \hat{J} = \begin{bmatrix}
1 - \frac{2 x_1}{k_p} -  \frac{y_1  \left(1+ \alpha \xi \right) } {\left(1+x_1+\alpha \xi\right)^2}  & \frac{- x_1}{1+x_1+\alpha \xi} & 0 & 0 \vspace{0.1cm}
  \\ 
\frac{\epsilon_1  \left(1 + \left(\alpha - 1\right) \xi\right) \ y_1}{(1+x_1+\alpha \xi)^2} &  \frac{\epsilon_1 \left(x_1 + \xi\right)}{1+x_1+\alpha \xi}  - q_1 - pc \xi y_1^{p-1} & 0 & 0
\vspace{0.1cm}
\\
0 & 0 & 1 - \frac{2 x_2}{k_c} - \frac{y_2}{(1+x_2)^2} & 
  \frac{- x_2}{1+x_2} \vspace{0.1cm}
  \\
  0 & q_2 & \frac{\epsilon_2 y_2}{(1+x_2)^2} & \frac{\epsilon_2 x_2}{1+x_2} - \delta_2 
  \end{bmatrix}
\label{general_jacobian_uni}
 \end{equation}
\subsubsection{Pest-free state in both \texorpdfstring{$\Omega_1 \ \& \ \Omega_2$}{Lg}\mbox{}}

\begin{lemma}
The equilibrium point $ \hat{E_1} = (0,y_1^*,0,y_2^*)$ exists if  $\xi > \frac{ q_1 }{\epsilon_1 - \alpha  q_1 }$ and $ \epsilon_1 > \alpha(\delta_1 + q_1) $. 
\label{lem:E1_existence_unidirectional}
  \end{lemma}
  \begin{proof}
See Appendix \eqref{proof_E1_existence_unidirectional}.
\end{proof}
\begin{lemma}
\label{lem:stability_pest_ext_drift}
The equilibrium point $ \hat{E_1} = (0,y_1^*,0,y_2^*)$ is locally asymptotically stable when, $y_1^*>\text{max} \left\{ \frac{\delta_2}{q_2}, 1+\alpha \xi \right\}$. 
\end{lemma}

\begin{proof}
    See Appendix \eqref{proof:stability_pest_ext_drift}.
\end{proof}
\begin{figure}
\begin{subfigure}{.33\textwidth}
\includegraphics[width = 5.6 cm, height=5.39cm]{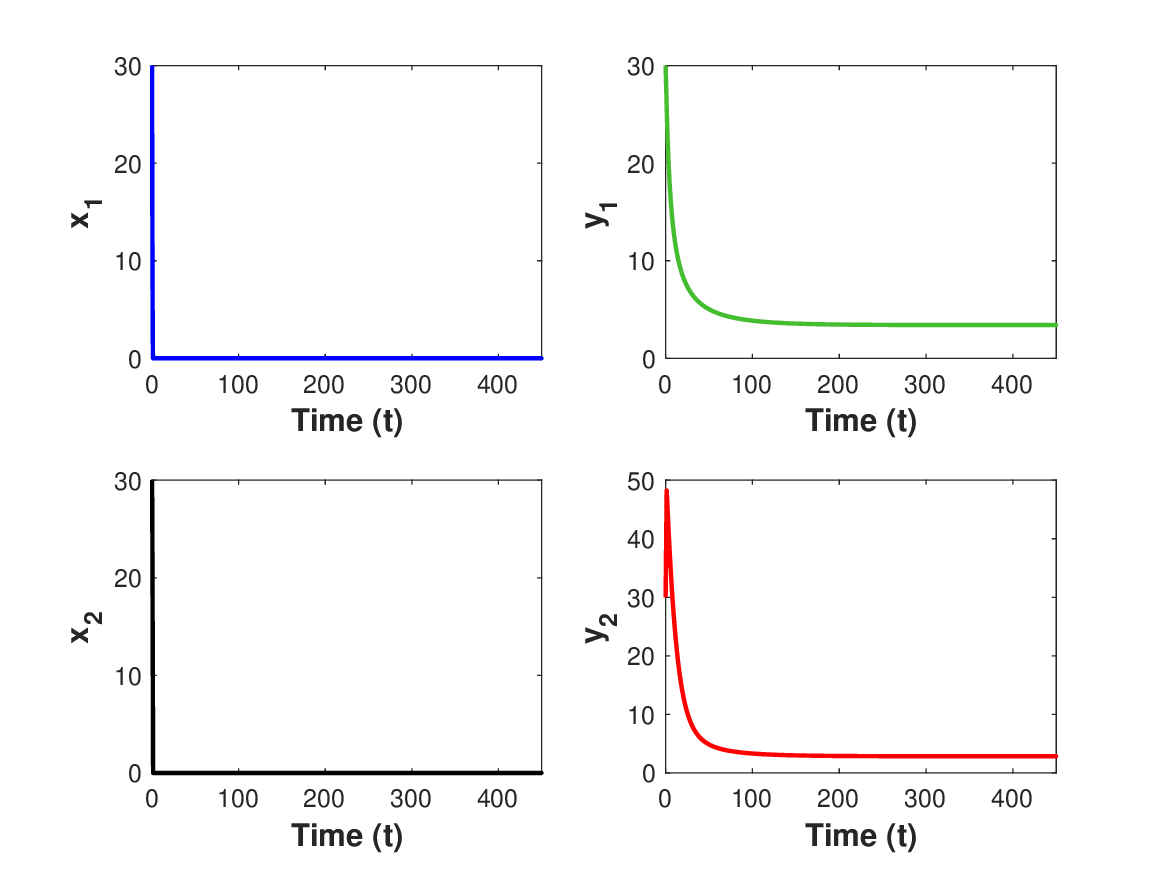}
\caption{}
\label{fig:x1_x2_extinction}
\end{subfigure}
\begin{subfigure}{.32\textwidth}
  \includegraphics[width= 5 cm, height=5cm]{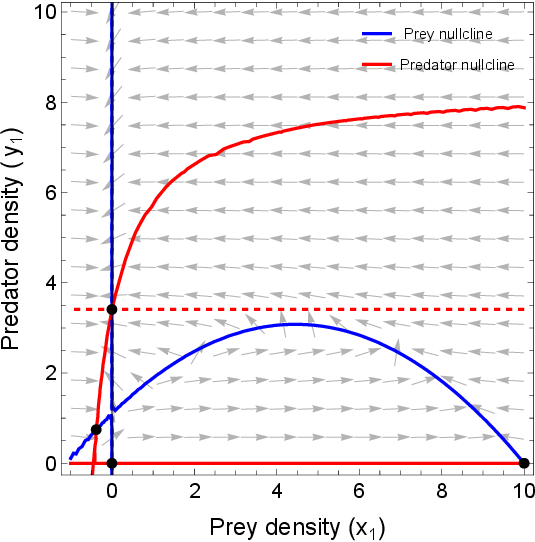}
  \caption{}
 \label{fig:patch1}
\end{subfigure} 
 \begin{subfigure}{.32\textwidth}
  \includegraphics[width= 5 cm, height=5cm]{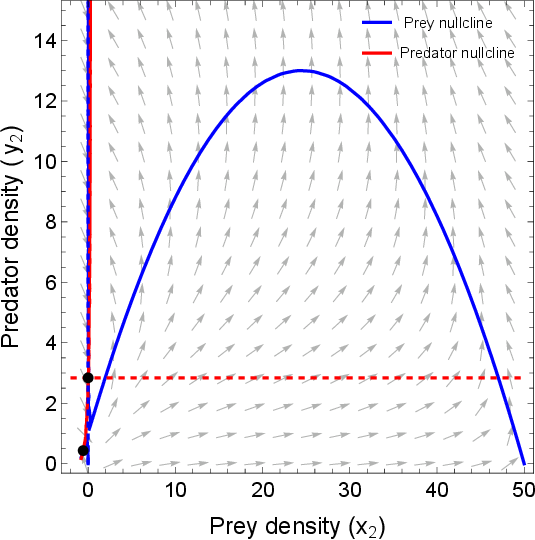}
  \caption{}
\label{fig:patch2}
\end{subfigure}
\caption{\textbf{Drift: Pest extinction in both patches.} Fig. \ref{fig:x1_x2_extinction} is the time series plot showing pest extinction in both patches $\Omega_1$ and $\Omega_2$ while predator populations reach an equilibrium level for the drift model \eqref{eq:model_drift_cann}. Figs. \ref{fig:patch1} and \ref{fig:patch2} represent the prey and predator nullclines in patch $\Omega_1$ and $\Omega_2$, respectively. The parameters used are 
$k_p=10,k_c=50,\alpha=0.1, \xi=0.99, \epsilon_1 = 0.3, \epsilon_2 =0.5,  \delta_2=0.12, q_1=0.25, q_2=0.1, c=0.006,p=2 \ \text{with I.C.,} \  x_1(0)=30,y_1(0)=30,x_2(0)=30,y_2(0)=30 $. }
\label{nullcline_complete_extinction}
 \end{figure}
 \subsubsection{Pest-free state  only in \texorpdfstring{$\Omega_2$}{Lg}\mbox{}}

 \begin{lemma}
    The equilibrium point $ \hat{E_2} = (x_1^*,y_1^*,0,y_2^*)$ exists if $ \epsilon_1 \xi < A (1+\alpha \xi)$  and {$ \epsilon_1 >  A $ where $A=q_1 + c \xi  (y_1^*)^{p-1}$}.
\label{lem:E2_existence_unidirectional}
  \end{lemma}
  
  \begin{proof}
See Appendix \eqref{proof_E2_existence_unidirectional}.
\end{proof}

\begin{lemma}
\label{lem:E2_stability_unidirectional}
The equilibrium point $ \hat{E_2} = (x_1^*,y_1^*,0,y_2^*)$ is conditionally locally asymptotically stable. 
\end{lemma}
\begin{proof}
See Appendix \eqref{proof_E2_stability_unidirectional}.
\end{proof}
\begin{figure}
 \begin{subfigure}{.33\textwidth}
\includegraphics[width = 5.6 cm, height=5.39cm]
{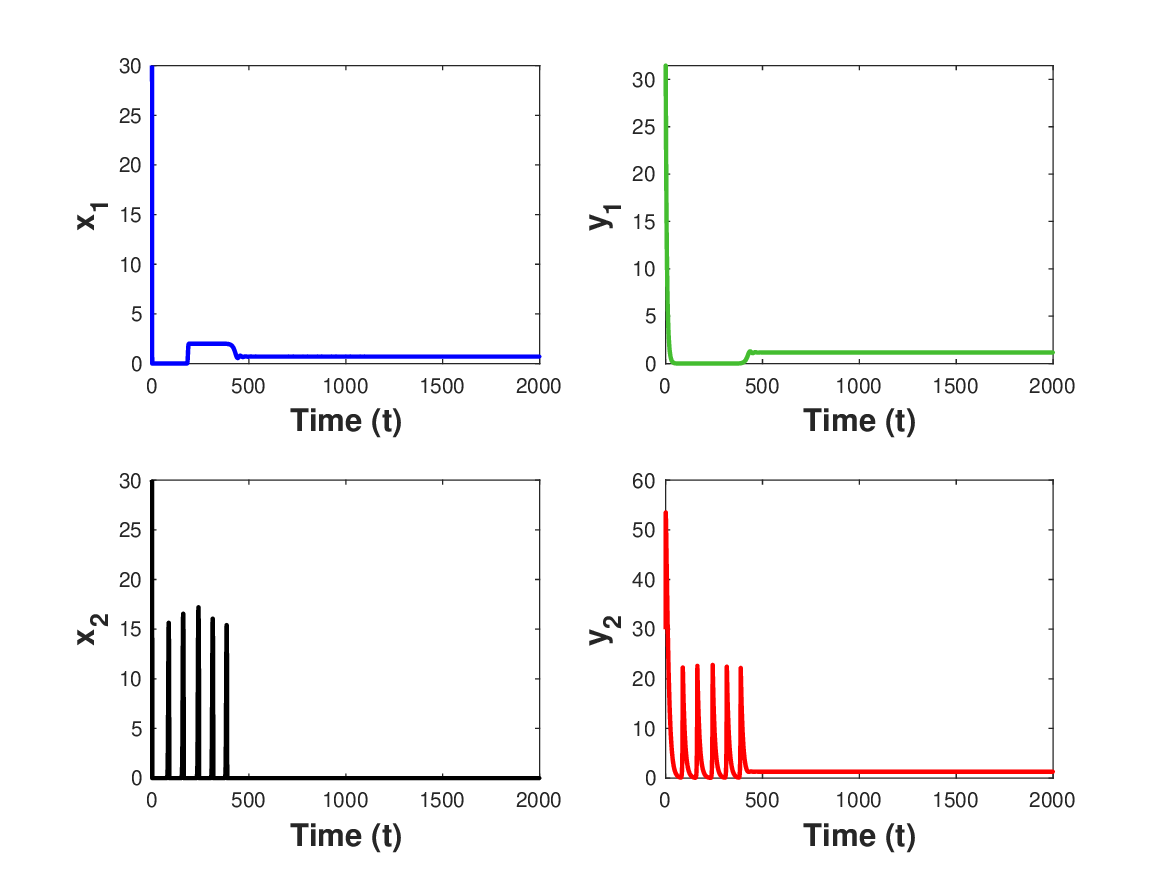}
\caption{}
\label{fig:x2_extinction_drift}
\end{subfigure}
   \begin{subfigure}{.32\textwidth}
  \includegraphics[width= 5cm, height=5cm]{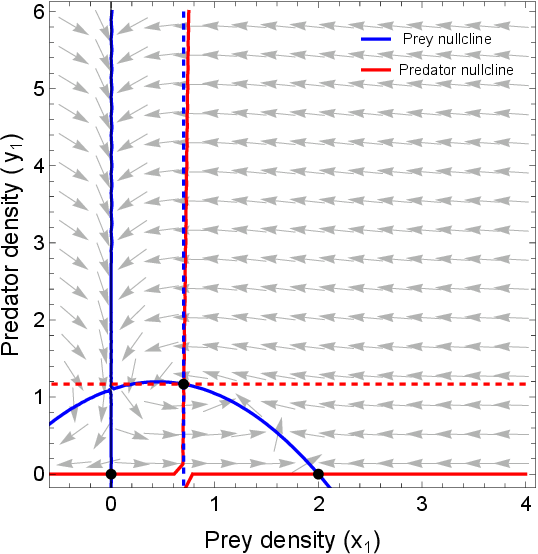}
\caption{}
 \label{fig:null_x2_0_patch1}
  \end{subfigure}
  \begin{subfigure}{.32\textwidth}
  \includegraphics[width= 5 cm, height=5cm]{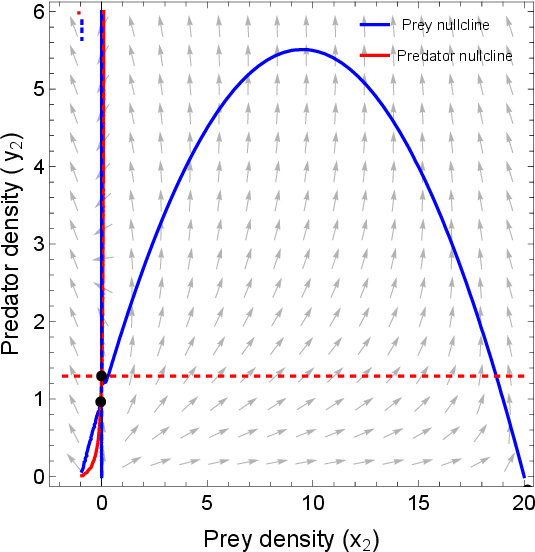}
\caption{}
\label{fig:null_x2_0_patch2}
 \end{subfigure}
 \caption{\textbf{Drift: Pest extinction only in the crop field patch.} Fig. \ref{fig:x2_extinction_drift} is the time series plot showing the pest extinction in the crop field patch $\Omega_2$ after oscillations and the populations in  $\Omega_1$ reaching an equilibrium level for the drift model \eqref{eq:model_drift_cann}. Figs. \ref{fig:null_x2_0_patch1} and \ref{fig:null_x2_0_patch2} represent the prey and predator nullclines in patch $\Omega_1$ and $\Omega_2$, respectively. The parameters used are 
$k_p=2,k_c=20,\alpha=0.48, \xi=0.2, \epsilon_1 = 0.45, \epsilon_2 =0.8, \delta_2=0.09, q_1=0.224, q_2=0.1, c=0.006,p=2 \ \text{with I.C.,} \  x_1(0)=30,y_1(0)=30,x_2(0)=30,y_2(0)=30 $.}
\label{x2_extinction_cann_nullcline_drift}
 \end{figure}
\subsection{Global stability of the equilibrium point \texorpdfstring{$\hat{E_1} = (0,y_1^*,0,y_2^*)$}{E1 = (0,y1*,0,y2*)}}
\begin{rem}
 The two-patch model \eqref{eq:model_drift_cann} can be considered as two distinct patches, as the dynamics in $\Omega_1$ are not affected by populations in $\Omega_2$, and when the predator population in $\Omega_1$ reaches an equilibrium level, it can be considered as a constant predator stocking in $\Omega_2$. 
\end{rem}

\begin{rem}
Fig. \ref{nullcline_complete_extinction} shows the time series plot of model \eqref{eq:model_drift_cann} where pest extinction occurs in both $\Omega_1$ and $\Omega_2$, and the nullclines plot of both the patches, showing that the prey and predator nullclines do not intersect with each other, and gives the pest extinction state.  
\end{rem}
 \subsubsection{Pest-free state in the patch \texorpdfstring {$\Omega_1$}{Lg}}
 \begin{thm}
Consider the following parametric restriction  on $c, c < c^*= min  \left\{ c_1^*,c_2^*,c_3^*,c_4^* \right\} $ where,
 $ c_1^*= \left( \frac{\epsilon_1 \xi-  q_1(1+\alpha \xi) }{\xi(1+\alpha \xi)^p}\right)^{\frac{1}{p-1}}, \quad 
c_2^* = \left(\frac{q_2}{\delta_2}\right)^{p-1}\left( \frac{\epsilon_1 \xi-  q_1(1+\alpha \xi) }{\xi(1+\alpha \xi)}\right), $
$c_3^*=
\left( \frac{\epsilon_1-q_1}{\xi}\right) \left(\frac{4 k_p}{(1+\alpha \xi +k_p)^2}\right)^{p-1} \ \text{and}, \
c_4^*= \frac{{h(x)}^{2-p}}{\xi} \left(\frac{4 \epsilon_1  (1+ \alpha \xi- \xi) }{(p-1)(1+k_p+\alpha \xi)^2 }\right)^{p-1}\left(\frac{k_p  }{ k_p-(1+\alpha \xi)}\right)^{p-1},
$
then  the equilibrium point $  (0,y_1^*)$  is globally stable. 
\label{lem:E1_global_unidirectional}
   \end{thm}
  \begin{proof}
\label{thm:proof_E1_global_unidirectional}
From Lemma \ref{lem:stability_pest_ext_drift}, we should have $y_1^*>\text{max} \left\{ \frac{\delta_2}{q_2}, 1+\alpha \xi \right\}$ else this equilibrium point will be unstable. This will hold if the following two conditions are also satisfy in terms of the parameter $ c$. 
\begin{equation}
c < \left( \dfrac{\epsilon_1 \xi-  q_1(1+\alpha \xi) }{\xi(1+\alpha \xi)^p}\right)^{\frac{1}{p-1}}
\label{c_first_drift}
\end{equation}

\begin{equation}
c < \left(\dfrac{q_2}{\delta_2}\right)^{p-1}\left( \dfrac{\epsilon_1 \xi-  q_1(1+\alpha \xi) }{\xi(1+\alpha \xi)}\right)
\label{c_second_drift}
\end{equation}

The equations representing dynamics in patch $\Omega_1$ for \eqref{eq:model_drift_cann} are given by:

  \begin{equation}
  \begin{aligned}
 \Dot{x_1}& = x_1 \left (1-\frac{x_1}{k_p}\right )-\frac{x_1 y_1}{1+x_1+\alpha \xi} \\
 \Dot{y_1}& = \epsilon_1 \left ( \frac{x_1+\xi}{1+x_1+\alpha \xi} \right)\ y_1  - q_1 y_1 -c \xi y_1^p  
 \end{aligned}
   \end{equation}
   
We also need that the maximum of the prey $(x_1)$ nullcline is below the horizontal asymptote of the predator $(y_1)$  nullcline, which requires the following inequality to hold, 
\begin{equation}
c< \left( \dfrac{\epsilon_1-q_1}{\xi}\right) \left(\dfrac{4 k_p}{(1+\alpha \xi +k_p)^2}\right)^{p-1}
    \label{c_third_drift}
\end{equation}

Now, in order to avoid the existence of an interior equilibrium point, the prey and predator nullclines should not intersect. Thus, we require that the predator nullcline $g(x_1)$ remain higher than the prey nullcline $f(x_1)$ on the interval $x_1 \in [0, \frac{1}{2}(k_p-1-\alpha \xi)]$. A sufficient condition for this is $g(0) \geq f(0)$ and min $g'(x_1) $ $> $  max $f'(x_1)$ for $x_1 \in [0, \frac{1}{2}(k_p-1-\alpha \xi)]$. The minimum of $g'(x_1)$ occurs at $x_1= \frac{1}{2}(k_p-1-\alpha \xi)$, and the maximum of $f'(x_1)$ occurs at $x_1=0$. Then, using these values and finding the inequality that satisfies the above condition is, 
\begin{equation}
c < \dfrac{{h(x)}^{2-p}}{\xi} \left(\dfrac{4 \epsilon_1  (1+ \alpha \xi- \xi) }{(p-1)(1+k_p+\alpha \xi)^2 }\right)^{p-1}\left(\dfrac{k_p  }{ k_p-(1+\alpha \xi)}\right)^{p-1}
    \label{c_fourth_drift}
\end{equation}
where $h(x) = \epsilon_1 \left ( \frac{x_1+\xi}{1+x_1+\alpha \xi} \right)  - q_1  $.
Thus, for no interior equilibrium to exist, we require taking the minimum of \eqref{c_first_drift}, \eqref{c_second_drift}, \eqref{c_third_drift}, and \eqref{c_fourth_drift}. Thus, only boundary $(k_p,0)$ and trivial $(0,0)$ equilibria exist, which are saddle and unstable, respectively. Thus, no periodic orbit exists.    
\end{proof}
 \subsubsection{Pest-free state in the patch \texorpdfstring {$\Omega_2$}{Lg}}

The equations representing dynamics in patch $\Omega_2$ are given by:
\begin{equation} \label{eq:patch_model_patch2_uni}
\begin{aligned}
 \Dot{x_2}& = x_2 \left (1- \frac{x_2}{k_c} \right)-\frac{x_2  y_2}{1+x_2} \\
\Dot{y_2}& = \epsilon_2 \left ( \frac{x_2}{1+x_2} \right ) \ y_2 - \delta_2 y_2 + q_2 y_1  
\end{aligned}
\end{equation}
Using the value of $y_1^*$ from \eqref{explicit_y1_comp_ext} we have, 
\begin{equation} \label{eq:patch_model_patch2_reduced_uni}
\begin{aligned}
 \Dot{x_2}& = x_2 \left (1- \frac{x_2}{k_c} \right)-\frac{x_2  y_2}{1+x_2} = x_2 \tilde f(x_2,y_2) \\
\Dot{y_2}& = \epsilon_2 \left ( \frac{x_2}{1+x_2} \right ) \ y_2 - \delta_2 y_2 - G_1 = y_2 \tilde g(x_2,y_2)- G_1
\end{aligned}
\end{equation}
where $G_1$ represents the constant predator stocking \citep{brauer1981constant}.
We state the following lemma,
\begin{thm}
\label{lem:pest_extinction_crop_field}
Consider the system \eqref{eq:patch_model_patch2_reduced_uni}. Then under the parametric restriction $\epsilon_2 > \delta_2\left(1-\frac{1}{k_c}\right)$ \ and , $\delta_2 < q_2 \left( \frac{\epsilon_1 \xi-  q_1(1+\alpha \xi) }{c\xi(1+\alpha \xi)}\right)^{\frac{1}{p-1}},$  we have that the pest free state $(0,y^{*}_{2})$ in the crop field to 
\eqref{eq:patch_model_patch2_reduced_uni}, is globally attracting for any positive $(x_{2}(0),y_{2}(0))$. 
\end{thm}
\begin{proof}
From the prey nullcline in patch $\Omega_2$ we have, 
\begin{equation}
    y_2^*  =  \left (1- \frac{x_2^*}{k_c} \right) \left(1+x_2^*\right) 
    \label{y2_x2}
\end{equation}
and from the predator nullcline in patch $\Omega_2$ we have, 
\begin{equation*}
  y_2^*  \ \tilde g(x_2^*,y_2^*) =  G_1 \implies y_2^* \left(   \epsilon_2 \left ( \frac{x_2^*}{1+x_2^*} \right ) - \delta_2 \right) =  G_1 = -q_2 y_1^*
  \end{equation*}
The constant predator stocking in terms of parameters can be written as, 
\begin{equation}
    G_1= -q_2 y_1^*= -q_2  \left( \dfrac{\epsilon_1 \xi-  q_1(1+\alpha \xi) }{c\xi(1+\alpha \xi)}\right)^{\frac{1}{p-1}}, \ p \neq 1
\end{equation}
The critical stocking rate for which the equilibrium reaches the pest-free state is given by \citep{brauer1981constant},

\begin{equation*}
  -G_{critical} = -\tilde g(0,1) \implies G_{critical} = -\delta_2 
\end{equation*}
Now, the pest extinction state in the crop field $\Omega_2$ is possible for all initial conditions if,

\begin{equation*}
\setlength{\jot}{10pt}
\begin{aligned}
&\epsilon_2 > \delta_2\left(1-\dfrac{1}{k_c}\right) \  \text{and,} \ G_1 < G_{critical} \implies
  y_2^* > 1\\
& \implies -q_2 y_1^* < -\delta_2 \implies q_2 y_1^* > \delta_2 \\
& \implies \delta_2 < q_2 \left( \dfrac{\epsilon_1 \xi-  q_1(1+\alpha \xi) }{c\xi(1+\alpha \xi)}\right)^{\frac{1}{p-1}}
\end{aligned}
\end{equation*}
This proves the lemma.
\end{proof}

\begin{thm}
\text{Consider the following parametric restrictions},
$ c < c^*= min \left\{ c_1^*,c_2^*,c_3^*,c_4^* \right\}$\text{,} $ \epsilon_2 > \delta_2\left(1-\frac{1}{k_c}\right) $ \text{and,} $ \delta_2 < q_2 \left( \frac{\epsilon_1 \xi-  q_1(1+\alpha \xi) }{c\xi(1+\alpha \xi)}\right)^{\frac{1}{p-1}} \ \text{where,}$
$ c_1^*= \left( \frac{\epsilon_1 \xi-  q_1(1+\alpha \xi) }{\xi(1+\alpha \xi)^p}\right)^{\frac{1}{p-1}}, \quad 
c_2^* = \left(\frac{q_2}{\delta_2}\right)^{p-1}\left( \frac{\epsilon_1 \xi-  q_1(1+\alpha \xi) }{\xi(1+\alpha \xi)}\right), 
$ 
$c_3^*=
\left( \frac{\epsilon_1-q_1}{\xi}\right) \left(\frac{4 k_p}{(1+\alpha \xi +k_p)^2}\right)^{p-1} \ \text{and}, \
c_4^*= \frac{{h(x)}^{2-p}}{\xi} \left(\frac{4 \epsilon_1  (1+ \alpha \xi- \xi) }{(p-1)(1+k_p+\alpha \xi)^2 }\right)^{p-1}\left(\frac{k_p  }{ k_p-(1+\alpha \xi)}\right)^{p-1}$
then the equilibrium point $ \hat{E_1} = (0,y_1^*,0,y_2^*)$  is globally stable. 
\label{thm:E1_global_unidirectional}
\end{thm}
\begin{proof}
The proof follows from Theorem
\ref{lem:E1_global_unidirectional} and \ref{lem:pest_extinction_crop_field}. 
    \label{proof_comp_pest_ext_uni}
\end{proof}
\subsection{Global stability of the equilibrium point  \texorpdfstring{$\hat{E_2} = (x_1^*,y_1^*,0,y_2^*)$}{E2 = (x1*,y1*,0,y2*)}}

 \begin{rem}
Considering the two-patch model \eqref{eq:model_drift_cann} as two distinct patches, we make use of the Dulac criterion to exclude the possibility of periodic orbits and establish the interior equilibrium as globally stable in $\Omega_1$, and the pest extinction state in $\Omega_2$ is proved globally stable via constant predator stocking. 
\end{rem}

\begin{rem}
Fig. \ref{x2_extinction_cann_nullcline_drift} shows the time series plot of model \eqref{eq:model_drift_cann}, where the dynamics in $\Omega_1$ exhibit coexistence and the pest extinction state is achieved in $\Omega_2$. The nullclines plot of both patches shows that the prey and predator nullclines intersect, yielding a stable interior equilibrium in patch $\Omega_1$, and in patch $\Omega_2$, the nullclines do not intersect, giving rise to the pest extinction state.    
\end{rem}
\subsubsection{Coexistence state in the patch \texorpdfstring {$\Omega_1$}{Lg}} 
\begin{thm}
Consider the parametric restriction $\xi > \frac{ q_1}{\epsilon_1- \alpha q_1}$, $
c > \left( \frac{\epsilon_1 \xi-  q_1(1+\alpha \xi) }{\xi(1+\alpha \xi)^p}\right)^{\frac{1}{p-1}}
$  and let $D$ be the region defined as,
$D=\left\{ (x_1, y_1) \in \mathbb{R}_+^2 \;\middle|\; 0 < x_1 < k_p,\; 0 < y_1 <y_{min}  \right\} \ \text{and if} \ 1<p<2$
where 
$y_{min} = \text{min}\left\{\left(\frac{\epsilon_1-q_1}{c \xi}\right)^\frac{1}{p-1},  \left(\frac{1}{2\xi c(1+\alpha \xi)(p-1)}\right)^\frac{1}{p-2}\right\} $ then, only one interior equilibrium exists and that is globally attracting. When $p=2$, then, for the interior equilibrium to be globally attracting, we require the following parametric restriction to hold,  $c > \frac{1}{2 \xi (1+\alpha\xi)}$.  
       \label{e2_patch1}
\end{thm}

\begin{proof}
In order for the interior equilibrium to exist, we have the following condition from Lemma \ref{lem:E2_existence_unidirectional}. 
$$0 < x_1 < k_p, \quad 0<y_1<\left(\frac{\epsilon_1-q_1}{c \xi}\right)^\frac{1}{p-1}$$
We will now use the Dulac criterion to exclude the existence of periodic orbits.
Consider the auxiliary function
$\phi(x_1,y_1) = \dfrac{1}{x_1 y_1}$,
\begin{equation*}
\setlength{\jot}{10pt}
\begin{aligned}
& \nabla \cdot \left( \phi(x, y) \frac{dx}{dt}, \, \phi(x, y) \frac{dy}{dt} \right)\\
&= \frac{\partial}{\partial x_1} \left( \frac{1}{x_1 y_1} \left( x_1 - \frac{x_1^2}{k_p} - \frac{x_1 y_1}{1+x_1 + \alpha \xi } \right) \right)\\
&+ \frac{\partial}{\partial y_1} \left( \frac{1}{x_1 y_1} \left( \epsilon_1 \left ( \frac{x_1+\xi}{1+x_1+\alpha \xi} \right)\ y_1  - q_1 y_1 -c \xi y_1^p  \right)  \right)\\
& = \frac{-1}{k_p y_1} 
+   \dfrac{1}{(1+ x_1+\alpha \xi)^2}  -  \dfrac{c \xi (p-1) y_1^{p-2}}{x_1}\\
& \leq   \dfrac{1}{(1+ x_1+\alpha \xi)^2}  -   \dfrac{c \xi(p-1)y_1^{p-2}}{x_1} \leq \dfrac{1}{2(1+\alpha \xi)x_1}  -   \dfrac{c \xi(p-1)y_1^{p-2}}{x_1}\\
& =   \dfrac{1}{ x_1}  \left(  \dfrac{1}{2(1+\alpha \xi)}  -   c \xi(p-1)y_1^{p-2}\right)
\end{aligned}
\end{equation*}
Thus, we require
\begin{equation*}
  y_1^{p-2}  > \dfrac{1}{2\xi c(1+\alpha \xi)(p-1)}     
    \end{equation*}
and since $p-2<0$ if $p \neq 2$ we get an upper bound on $y_1$,
\begin{equation*}
y_1  < \left(\dfrac{1}{2\xi c(1+\alpha \xi)(p-1)}\right)^\frac{1}{p-2}     
\end{equation*}
Now, for the case when $p=2$, the interior equilibrium is globally attracting if the following parametric restriction is satisfied,  
$ c > \frac{1}{2 \xi (1+\alpha\xi)}$.
Thus, via the Dulac criterion, the limit cycles would not exist.
Looking at the patch $\Omega_1$, the extinction state $(0,0)$ is an unstable node and the predator-free state $(k_p,0)$ is a saddle, based on the parametric restriction mentioned on $\xi$. Also, the pest extinction state $(0,y^*)$ will be a saddle if \ $
c > \left( \frac{\epsilon_1 \xi-  q_1(1+\alpha \xi) }{\xi(1+\alpha \xi)^p}\right)^{\frac{1}{p-1}}
$, see Lemma \ref{lem:stability_pest_ext_drift}. Therefore, the only interior equilibrium that exists is globally attracting.
\label{e2_patch1_proof}
\end{proof}
 \subsubsection{Pest-free state in the patch \texorpdfstring {$\Omega_2$}{Lg}}
From equation \eqref{eq:patch_model_patch2_uni} and using the value of $y_1^*$ from \eqref{x2_0_eq_3_uni} we have, 
\begin{equation} \label{eq:patch_model_patch2_reduced_uni2}
\begin{aligned}
 \Dot{x_2}& = x_2 \left (1- \frac{x_2}{k_c} \right)-\frac{x_2  y_2}{1+x_2} = x_2 \tilde f(x_2,y_2) \\
\Dot{y_2}& = \epsilon_2 \left ( \frac{x_2}{1+x_2} \right ) \ y_2 - \delta_2 y_2 - G_2 = y_2 \tilde g(x_2,y_2)- G_2
\end{aligned}
\end{equation}

\begin{thm}
Consider the system \eqref{eq:patch_model_patch2_reduced_uni2}. Then under the parametric restriction $\epsilon_2 > \delta_2\left(1-\frac{1}{k_c}\right)$ \ and , $\delta_2 < q_2 \left(\frac{\epsilon_1 \left ( \frac{x_1^* +\xi}{1+x_1^* +\alpha \xi} \right)  - q_1} { c \xi}\right)^{\frac{1}{p-1}}$, we have that the pest free state $(0,y^{*}_{2})$ in the crop field to 
\eqref{eq:patch_model_patch2_reduced_uni}, is globally attracting for any positive $(x_{2}(0),y_{2}(0))$. 
\label{e2_patch2}
\end{thm}

\begin{proof}
From the predator nullcline in patch $\Omega_2$, we have, 
\begin{equation*}
  y_2^*  \ \tilde g(x_2^*,y_2^*) =  G_2 \implies y_2^* \left(   \epsilon_2 \left ( \frac{x_2^*}{1+x_2^*} \right ) - \delta_2 \right) =  G_2 = -q_2 y_1^*
  \end{equation*}
From equation \eqref{eq:patch_model_patch2_reduced_uni} we know that the system can be written in the form of a constant predator stocking represented by $G_2$. The expression of $y_1^*$ is written in terms of $x_1^*$ because finding an explicit expression in terms of parameters is not feasible due to non nonlinearity involved. The constant predator stocking is then given by, 
\begin{equation}
G_2= -q_2 y_1^*= -q_2 \left(\dfrac{\epsilon_1 \left ( \frac{x_1^* +\xi}{1+x_1^* +\alpha \xi} \right)  - q_1} { c \xi}\right)^{\frac{1}{p-1}} , \ p \neq 1
\end{equation}
The critical stocking rate for which the equilibrium reaches the pest-free state is given by \citep{brauer1981constant},
\begin{equation*}
  -G_{critical} = -\tilde g(0,1) \implies G_{critical} = -\delta_2 
\end{equation*}
Now, the pest free state in the crop field patch $\Omega_2$ is possible for all initial conditions if,
\begin{equation*}
\setlength{\jot}{10pt}
\begin{aligned}
&\epsilon_2 > \delta_2\left(1-\dfrac{1}{k_c}\right) \  \text{and,} \ G_2 < G_{critical} \implies
y_2^* > 1\\
& \implies -q_2 y_1^* < -\delta_2 \implies q_2 y_1^* > \delta_2 \\
& \implies \delta_2 < q_2 \left(\dfrac{\epsilon_1 \left ( \frac{x_1^* +\xi}{1+x_1^* +\alpha \xi} \right)  - q_1} { c \xi}\right)^{\frac{1}{p-1}} 
\end{aligned}
\end{equation*}
This proves the lemma.
\label{e2_patch2_proof}
\end{proof}
\begin{thm}
Let $D$ be the region, $D=\left\{ (x_1, y_1) \in \mathbb{R}_+^2 \;\middle|\; 0 < x_1 < k_p,\; 0 < y_1 <y_{min}  \right\} \ $ and, consider the parametric restrictions in patch $\Omega_1$, $\xi > \frac{ q_1}{\epsilon_1- \alpha q_1}$, $
c > \left( \frac{\epsilon_1 \xi-  q_1(1+\alpha \xi) }{\xi(1+\alpha \xi)^p}\right)^{\frac{1}{p-1}}$
\text{and if} \ $ p \in (1,2)$
where  $y_{min} = \text{min}\left\{\left(\frac{\epsilon_1-q_1}{c \xi}\right)^\frac{1}{p-1},  \left(\frac{1}{2\xi c(1+\alpha \xi)(p-1)}\right)^\frac{1}{p-2}\right\} $, when $p=2$, we need $c > \frac{1}{2 \xi (1+\alpha\xi)}$ and the following parametric restriction in patch $\Omega_2$, $\epsilon_2 > \delta_2\left(1-\frac{1}{k_c}\right)$ \ and , $\delta_2 < q_2 \left(\frac{\epsilon_1 \left ( \frac{x_1^* +\xi}{1+x_1^* +\alpha \xi} \right)  - q_1} { c \xi}\right)^{\frac{1}{p-1}}$, then the equilibrium point $ \hat{E_2} = (x_1^*,y_1^*,0,y_2^*)$ is  globally stable.
\label{thm:E2_global_unidirectional}
  \end{thm}
\begin{proof}
The proof follows from Theorem \ref{e2_patch1} and \ref{e2_patch2}.
\label{thm:proof_E2_global_unidirectional}
\end{proof}

\section{Modeling Dispersal Between \texorpdfstring{$\Omega_1$ \& $\Omega_2$}{Omega1 and Omega2}}
\label{dispersal_section}

 \begin{equation} \label{eq:patch_model2}
\begin{aligned}
\Dot{x_1}& = x_1 \left (1-\frac{x_1}{k_p}\right )-\frac{x_1 y_1}{1+x_1+\alpha \xi} \\
 \Dot{y_1}& = \epsilon_1 \left ( \frac{x_1+\xi}{1+x_1+\alpha \xi} \right)\ y_1  -c \xi y_1^p - q_1 y_1  +q_4 y_2, \   \quad 1< p \leq 2 \\
 \Dot{x_2}& = x_2 \left (1- \frac{x_2}{k_c} \right)-\frac{x_2  y_2}{1+x_2} \\
\Dot{y_2}& = \epsilon_2 \left ( \frac{x_2}{1+x_2} \right ) \ y_2 - \delta_2 y_2 + q_2 y_1 - q_3 y_2 
\end{aligned}
\end{equation}
Here, the assumptions on patches $\Omega_1 \ \& \ \Omega_2$ are the same as mentioned in Section \ref{drfit_model}. The predators now disperse between the two patches, where $q_1$ is the rate predators move out of the prairie strip ($\Omega_1$) after gaining energy from additional food and then enter the crop field ($\Omega_2$) at a rate of $q_2$. Similarly, the predators disperse out of the crop field since AF can act as an attractant, at a rate of $q_3$, and enter the strip with a rate of $q_4$. The assumption here on dispersal rates are \  $ q_2 \leq q_1 \ \text{and} \  q_4 \leq q_3$. Note, if $q_3=q_4=0$ then the model reduces to the drift model \eqref{eq:model_drift_cann}.  

\subsection{Mathematical Analysis}
\begin{thm}
Assume the parameters $k_p,k_c,\epsilon_1,\epsilon_2,\xi,\alpha$ are all positive and the dispersal rates $q_1,q_2,q_3,q_4$ are non-negative. Then, the model \eqref{eq:patch_model2} is positively invariant in $\mathbb{R}_+^4$. 
\label{positively invariant}
\end{thm}
\begin{proof}
See Appendix \eqref{proof:positve_dispersal}.
\end{proof}
We now consider the existence and local stability analysis of the biologically relevant equilibrium points for the system \eqref{eq:patch_model2}. The Jacobian matrix $(\tilde J)$ for the  additional food patch model \eqref{eq:patch_model2} is given by:
\begin{equation} 
\tilde{J} = \begin{bmatrix}
1 - \frac{2 x_1}{k_p} -  \frac{y_1  \left(1+ \alpha \xi \right) } {\left(1+x_1+\alpha \xi\right)^2}  & \frac{- x_1}{1+x_1+\alpha \xi} & 0 & 0 \vspace{0.1cm}
  \\ 
\frac{\epsilon_1  \left(1 + \left(\alpha - 1\right) \xi\right) \ y_1}{(1+x_1+\alpha \xi)^2} &  \frac{\epsilon_1 \left(x_1 + \xi\right)}{1+x_1+\alpha \xi} - c\xi p{y_1}^{p-1} - q_1 & 0 & q_4 
\vspace{0.1cm}
\\
0 & 0 & 1 - \frac{2 x_2}{k_c} - \frac{y_2}{(1+x_2)^2} & 
  \frac{- x_2}{1+x_2} \vspace{0.1cm}
  \\
  0 & q_2 & \frac{\epsilon_2 y_2}{(1+x_2)^2} & \frac{\epsilon_2 x_2}{1+x_2} - \delta_2 - q_3
  \end{bmatrix}
\label{general_jacobian_dispersal}
 \end{equation}
\subsubsection{Pest-free state in both \texorpdfstring{$\Omega_1 \ \& \ \Omega_2$}{Lg}\mbox{}}

\begin{lemma}
The equilibrium point $ \Tilde{E_1} = (0,y_1^*,0,y_2^*)$ exists if $\xi >  \frac{B}{\epsilon_1-\alpha B}, \ \text{and}, \ \epsilon_1-\alpha B>0 $ where $B=q_1 - \frac{q_4 q_2}{\delta_2 + q_3}$. 
\label{lem:E1_existence_dispersal}
  \end{lemma}
  \begin{proof}
  See Appendix \eqref{proof:E1_existence_dispersal}.
      \end{proof}
 \begin{lemma}
\label{lem:stability_pest_ext_dispersal}
The equilibrium point $ \Tilde{E_1} = (0,y_1^*,0,y_2^*)$ is locally asymptotically stable if,  $y_1^*>\text{max} \left\{ \frac{\delta_2+q_3}{q_2}, 1+\alpha \xi \right\}.$
\end{lemma}

\begin{proof}
See Appendix \eqref{proof:stability_pest_ext_dispersal}.
\end{proof}
 
\begin{figure}
 \begin{subfigure}{.33\textwidth}
\includegraphics[width = 5.6cm, height=5.39cm]{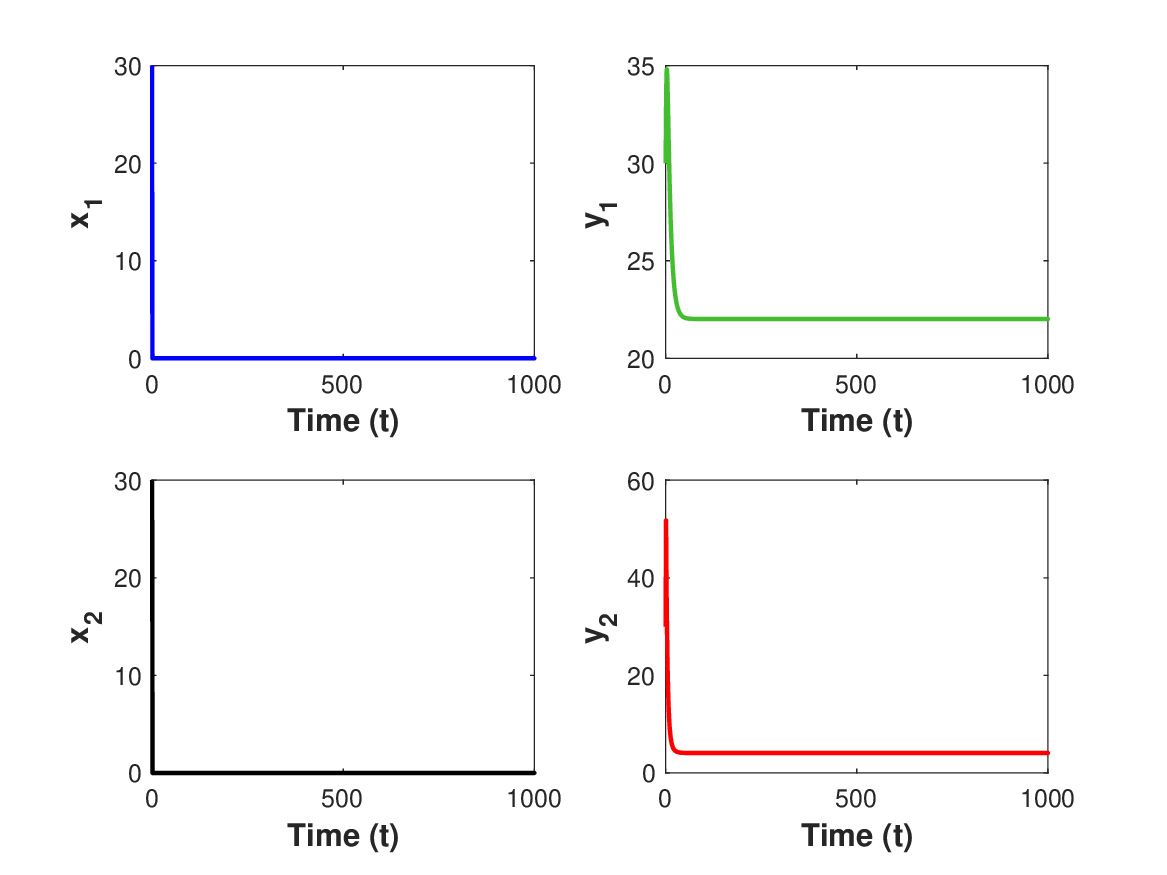}
\caption{}
\label{fig:x1_x2_ext_dispersal}
\end{subfigure}
  \begin{subfigure}{.32\textwidth}
  \includegraphics[width= 5.cm, height=5cm]{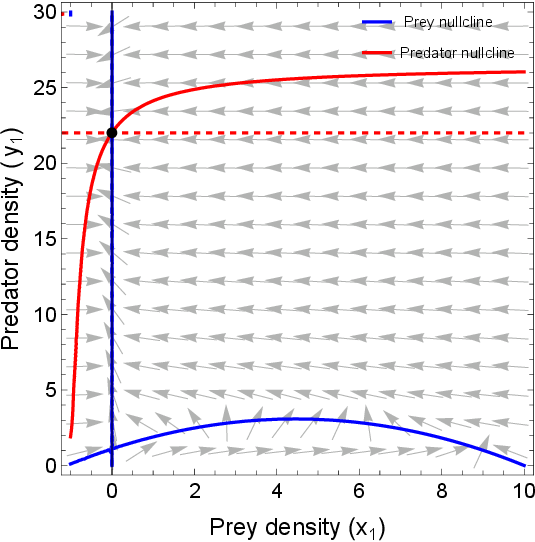}
  \caption{}
\label{fig:null_disp_comp_patch1}
  \end{subfigure}
  \begin{subfigure}{.32\textwidth}
  \includegraphics[width= 5.cm, height=5cm]{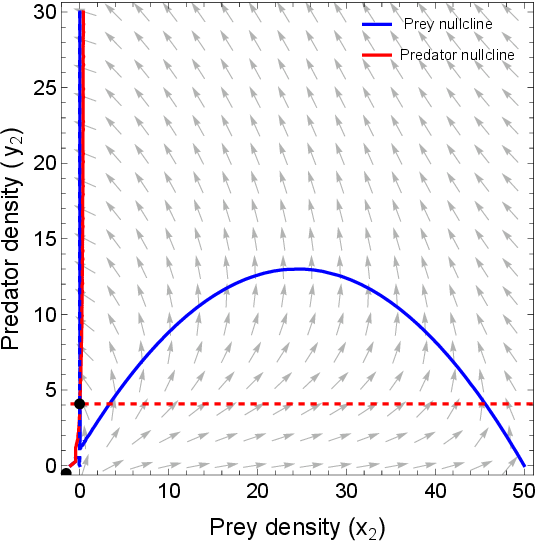}
  \caption{}
\label{fig:null_disp_comp_patch2}
\end{subfigure}
 \caption{\textbf{Dispersal: Pest extinction in both patches.} Fig. \ref{fig:x1_x2_ext_dispersal} is the time series plot showing pest extinction in both patches $\Omega_1$ and $\Omega_2$ while predator populations reach an equilibrium level for the dispersal model \eqref{eq:patch_model2}. Figs. \ref{fig:null_disp_comp_patch1} and \ref{fig:null_disp_comp_patch2} represent the prey and predator nullclines in patch $\Omega_1$ and $\Omega_2$, respectively.  The parameters used are 
$k_p=10,k_c=50,\alpha=0.1, \xi=0.99, \epsilon_1 = 0.3, \epsilon_2 =0.8, \delta_2=0.12, q_1=0.158, q_2=0.05,q_3=0.15,q_4=0.1, c=0.006,p=2 \ \text{with I.C.,} \  x_1(0)=30,y_1(0)=30,x_2(0)=30,y_2(0)=30 $. }
\label{com_ext_disp_nullcline}
 \end{figure}

     
 \subsubsection{Pest-free state only in \texorpdfstring{$\Omega_2$}{Lg}\mbox{}}

 \begin{lemma}
    The equilibrium point $ \Tilde{E_2} = (x_1^*,y_1^*,0,y_2^*)$ exists if if  $\epsilon_1 \xi < D(1+\alpha \xi)$ and {$ \epsilon_1 >  D $} where, $ D= c \xi  (y_1^*)^{p-1}+ q_1   -\frac{q_4 q_2} {\delta_2 + q_3}$.
\label{lem:E2_existence_dispersal}
  \end{lemma}
  
\begin{figure}
 \begin{subfigure}{.33\textwidth}
\includegraphics[width = 5.6cm, height=5.39cm]
{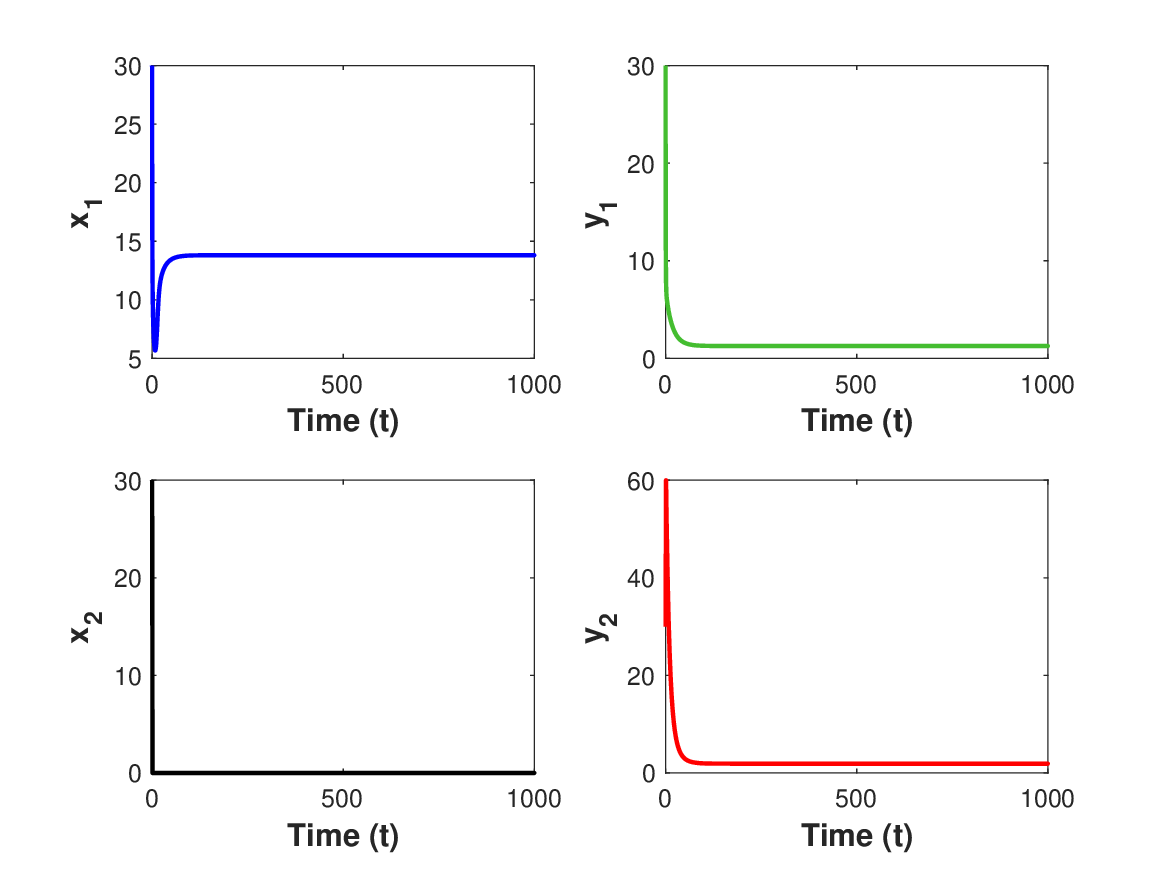}
\caption{}
\label{fig:x2_extinction_dispersal}
\end{subfigure}
  \begin{subfigure}{.32\textwidth}
  \includegraphics[width= 5 cm, height=5cm]{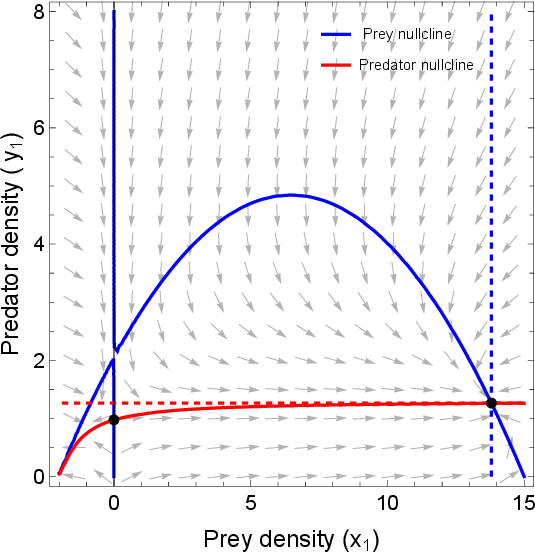}
  \caption{}
\label{fig:null_x2_0_patch1_dispersal}
 \end{subfigure}
   \begin{subfigure}{.32\textwidth}
  \includegraphics[width= 5 cm, height=5cm]{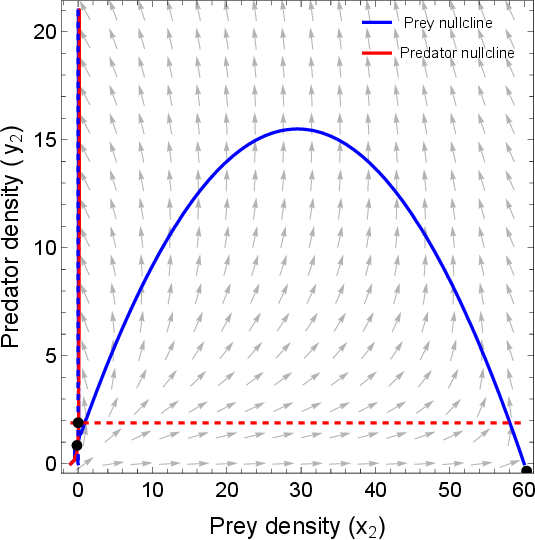}
  \caption{}
\label{fig:null_x2_0_patch2_dispersal}
 \end{subfigure}
 \caption{ \textbf{Dispersal: Pest extinction only in the crop field patch.} Fig. \ref{fig:x2_extinction_dispersal} is the time series plot showing the pest extinction in the crop field patch $\Omega_2$ and coexistence in the patch $\Omega_1$ for the dispersal model \eqref{eq:patch_model2}. Figs.  \ref{fig:null_x2_0_patch1_dispersal} and \ref{fig:null_x2_0_patch2_dispersal} represent the prey and predator nullclines in patch $\Omega_1$ and $\Omega_2$, respectively.  The parameters used are 
$k_p=15,k_c=60,\alpha=0.7, \xi=1.5, \epsilon_1 = 0.3, \epsilon_2 =0.8, \delta_2=0.02, q_1=0.258, q_2=0.18, q_3=0.1,q_4=0.08,c=0.08,p=2 \ \text{with I.C.,} \  x_1(0)=30,y_1(0)=30,x_2(0)=30,y_2(0)=30 $.}
\label{x2_extinction_cann_nullcline_dispersal}
 \end{figure}

  \begin{proof}
See Appendix \eqref{proof_E2_existence_dispersal}.
\end{proof} 
\begin{lemma}
\label{lem:E2_stability_dispersal}
The equilibrium point $ \Tilde{E_2} = (x_1^*,y_1^*,0,y_2^*)$ is conditionally locally asymptotically stable. 
\end{lemma}
\begin{proof}
See Appendix \eqref{proof_E2_stability_dispersal}.
\end{proof}
\begin{rem}
Fig. \ref{x2_extinction_cann_nullcline_dispersal} shows the time series plot of \eqref{eq:patch_model2}, where the dynamics in $\Omega_1$ exhibit coexistence and the pest extinction state is achieved in $\Omega_2$. The nullclines plot of both patches shows that the prey and predator nullclines intersect, yielding a stable interior equilibrium in patch $\Omega_1$, and in patch $\Omega_2$, the nullclines do not intersect, giving rise to the pest extinction state due to constant predator stocking.    
\end{rem}
 \subsection{Global stability of the equilibrium point \texorpdfstring{$\tilde{E_1} = (0,y_1^*,0,y_2^*)$}{E1 = (0,y1*,0,y2*)}}
\begin{figure}
\centering
\includegraphics[width=6cm]{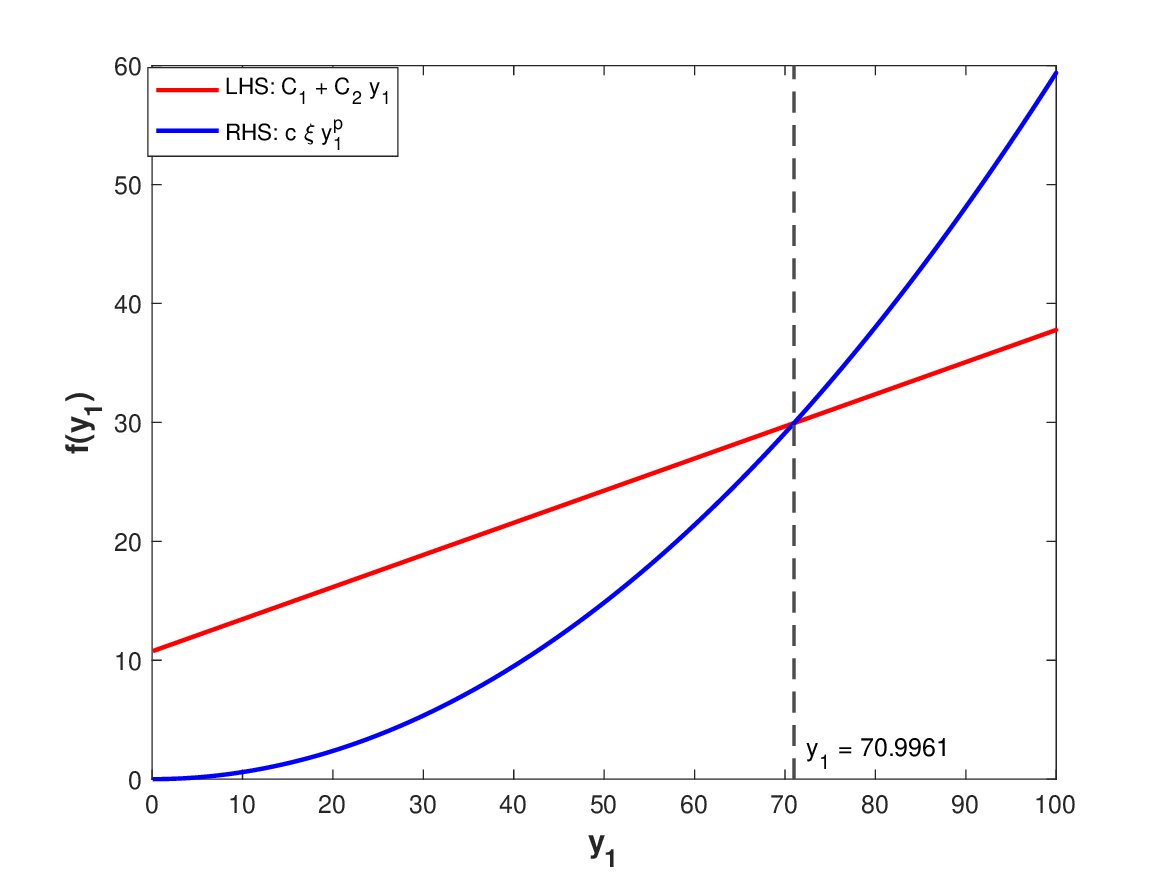}
\caption{This figure shows that beyond the threshold value $M$, which is represented by the black dotted line, the function $f(y_1)<0$ since the RHS (blue curve) grows faster than the LHS (red curve), making $\dot{L}\leq 0$ in $\Omega$. The parameters used are 
$k_p=10,k_c=50,\alpha=0.1, \xi=0.99, \epsilon_1 = 0.3, \epsilon_2 =0.8, \delta_2=0.12, q_1=0.158, q_2=0.05,q_3=0.15,q_4=0.1, c=0.006,p=2$.}
\label{fig:inequality_comp_dispersal}
\end{figure}

 \begin{rem}
 Fig. \ref{com_ext_disp_nullcline} shows the time series plot of model \eqref{eq:patch_model2} where pest extinction occurs in both $\Omega_1$ and $\Omega_2$, and the nullclines plot of both the patches, showing that the prey and predator nullclines do not intersect with each other, yielding the pest extinction states in both patches. 
    
 \end{rem}
\begin{thm}
If  $y_1^*>\text{max} \left\{ \frac{\delta_2+q_3}{q_2}, 1+\alpha \xi \right\}$ then, the equilibrium point $ \Tilde{E_1} = (0,y_1^*,0,y_2^*)$ is globally asymptotically stable.
\label{e1_patch1_dispersal} 
\end{thm}
\begin{proof}
Let $D$ be the region where, $D=\{(x_1,y_1,x_2,y_2) \in \mathbb{R}_+^4\, |\, x_1\geq0,\, y_1\geq0, \ x_2\geq0, y_2\geq0\}$. We define a function,
$L(x_1, x_2, y_1, y_2) = \epsilon_1 x_1 + \epsilon_2 x_2 + y_1 + y_2$, which is non-negative and  $L \rightarrow \infty$ as $||(x_1,y_1,x_2,y_2)|| \rightarrow \infty $, so radially unbounded. Upon  differentiating $L$ along the positive solutions of \eqref{eq:patch_model2}, 
\begin{equation*}
\begin{aligned}
\dot{L} &= \epsilon_1 \dot{x}_1 + \epsilon_2 \dot{x}_2 + \dot{y}_1 + \dot{y}_2 \\
\dot{L} &= \epsilon_1 x_1 \left(1 - \frac{x_1}{k_p} \right) - \epsilon_1 \frac{x_1 y_1}{1 + x_1 + \alpha \xi}
+ \epsilon_1 \left( \frac{x_1 + \xi}{1 + x_1 + \alpha \xi} \right) y_1 - c \xi y_1^p - q_1 y_1 + q_4 y_2 \\
&\quad + \epsilon_2 x_2 \left(1 - \frac{x_2}{k_c} \right) - \epsilon_2 \frac{x_2 y_2}{1 + x_2}
+ \epsilon_2 \left( \frac{x_2}{1 + x_2} \right) y_2 - \delta_2 y_2 + q_2 y_1 - q_3 y_2
\end{aligned}
\end{equation*}
Canceling the interaction terms involving $\epsilon_1$ and $\epsilon_2$ gives, 
\begin{equation}
\begin{aligned}
\dot{L} &= \epsilon_1 x_1 \left(1 - \frac{x_1}{k_p} \right) + \frac{ \epsilon_1 \xi y_1}{1 + x_1 + \alpha \xi}
- c \xi y_1^p - q_1 y_1 + q_4 y_2 \\
&\quad + \epsilon_2 x_2 \left(1 - \frac{x_2}{k_c} \right) - \delta_2 y_2 + q_2 y_1 - q_3 y_2
\end{aligned}
\label{simplified_xpr}
\end{equation}
Notice, $1+x_1+\alpha \xi \geq 1+\alpha \xi$ and from the assumptions on the dispersal rates in model \eqref{eq:patch_model2} we have, $q_2 \leq q_1$ and $q_4 \leq q_3$, this implies $- q_1 y_1 + q_2 y_1 \leq 0, \ q_4 y_2 - q_3 y_2 \leq 0$ \ so,  $(q_4 - q_3 -\delta_2 )y_2 \leq 0 $ then, 
\begin{equation*}
\dot{L} \leq \epsilon_1 x_1 \left(1 - \frac{x_1}{k_p} \right) +  \frac{\epsilon_1  \xi y_1}{1 + \alpha \xi}
- c \xi y_1^p + \epsilon_2 x_2 \left(1 - \frac{x_2}{k_c} \right) 
\end{equation*} 
Since the max for the logistic terms involving $x_1, x_2$ occurs at $\frac{k_p}{2}, \frac{k_c}{2}$  respectively. Also, for initial conditions $x_1(0),x_2(0) > 0$, the logistic dynamics ensures that $0 < x_1(t) < k_p$ and $0 < x_2(t) < k_c$, for all sufficiently large $t$, and any perturbation from the respective carrying capacities decays to zero as $t \to \infty$. Therefore, any deviation can be absorbed into a vanishing term $\epsilon(t) \to 0$. Then, we have the following bounds, 
\begin{equation*}
\dot{L} \leq  \frac{\epsilon_1 k_p}{4} + \frac{\epsilon_2 k_c}{4}   +  \epsilon(t) + \left(\frac{\epsilon_1  \xi }{1 + \alpha \xi}\right)y_1
- c \xi y_1^p,  \ \text{and as}  \ t \rightarrow  \infty, \epsilon(t) \rightarrow 0,
\end{equation*}
Neglecting the vanishing term $\epsilon(t)$ in the long-term limit gives,
\begin{equation*}
\dot{L} \leq  \frac{\epsilon_1 k_p}{4} + \frac{\epsilon_2 k_c}{4}   +  \left(\frac{\epsilon_1  \xi }{1 + \alpha \xi}\right)y_1
- c \xi y_1^p = f(y_1)
\end{equation*}
\begin{equation*}
 \text{Let,}  \  C_1 =  \frac{\epsilon_1 k_p}{4} + \frac{\epsilon_2 k_c}{4} \  \text{and} \  C_2= \frac{\epsilon_1  \xi }{1 + \alpha \xi}, \  \text{So,}  \ f(y_1)=C_1 +  C_2 y_1- c \xi y_1^p
\end{equation*}
We require, $f(y_1)\leq 0 \implies C_1 +  C_2 y_1 \leq c \xi y_1^p $. Since $1<p \leq 2$, the right-hand side grows faster than the linear terms on the left side. This inequality will hold for $y_1 \geq M$, where $M$ is given by the exact solution where $f(y_1)=0$. Thus, there exists a threshold value $M$ for which $f(y_1)<0$. So, the bounded subset of $D$ can be written as, $\Omega = \{(x_1,y_1,x_2,y_2) \in D\, |L(x_1, x_2, y_1, y_2) \le L_0,\ x_1 \leq k_p, \, y_1\geq M, \, x_2\leq k_c \}$, where $L_0 = L(x_1(0), x_2(0), y_1(0), y_2(0))$ and we have $\dot{L} \leq 0$ on this domain $\Omega$. This has been checked graphically as well, see Fig. \ref{fig:inequality_comp_dispersal}. After surpassing the threshold of $y_1$, we observed $f(y_1)<0$. Now, look at the set where $\dot{L} = 0$ so from \eqref{simplified_xpr} we have,
\begin{equation*}
\epsilon_1 x_1 \left(1 - \frac{x_1}{k_p} \right) + \frac{ \epsilon_1 \xi y_1}{1 + x_1 + \alpha \xi}
- c \xi y_1^p +(q_2-q_1) y_1 
 + \epsilon_2 x_2 \left(1 - \frac{x_2}{k_c} \right) - \delta_2 y_2 + (q_4 - q_3) y_2=0
\end{equation*}
Note that the points where $x_1=k_p, x_1=k_c$ are not invariant under the flow and can not belong to the omega limit set for $y_1,y_2 \neq 0$. So, in the invariant set, we only have $x_1=x_2=0$. And now, $y_1, y_2$ should satisfy the reduced system.
\begin{equation*}
\left(\frac{ \epsilon_1 \xi}{1 + \alpha \xi} - c \xi y_1^{p-1} +q_2-q_1\right) y_1 
=( q_3+ \delta_2-q_4) y_2
\end{equation*}
The above equation only holds at the equilibrium point values,  since from \eqref{x1_x2_0_eq1} using the value of $y_2^*$ and from \eqref{x1_x2_0_eq2} using the expression for $y_1^*$ we have,
\begin{equation*}
\left(\frac{ \epsilon_1 \xi}{1 + \alpha \xi} - c \xi (y_1^*)^{p-1}  +q_2-q_1\right) y_1^* 
=q_2 y_1^* -\dfrac{q_4 q_2}{\delta_2 + q_3} \ y_1^*
\end{equation*}
But we already know, 
\begin{equation*}
\frac{\epsilon_1 
 \xi}{1+\alpha \xi} \  - q_1  - c \xi (y_1^*)^{p-1} +\dfrac{q_4 q_2}{\delta_2 + q_3} = 0 
\end{equation*}
So, the largest invariant set $E$ where $\dot{V} =0$ is, 
\begin{equation*}
\begin{aligned}
& E = \left\{ (x_1, y_1, x_2, y_2) \in \Omega \;\middle|\; x_1 =0 ,\; y_1 = y_1^*, \; x_2 =0 ,\; y_2 = y_2^*  \right\}\\
& E = \left\{\tilde{E_1} = (0,y_1^*,0,y_2^*)  \right\}
\end{aligned}
\end{equation*}
\label{e1_patch1_proof_dispersal}
So, by LaSalle’s invariance principle, since $L$ is radially unbounded and $\dot{L} \leq 0$ in a positively invariant region $\Omega$, and the largest invariant set where   $\dot{L} = 0$ being the singleton equilibrium point $\tilde{E_1}$ so, all solution trajectories approach $\tilde{E_1}$. Thus, from Lemma \ref{lem:stability_pest_ext_dispersal}, if $y_1^*>\text{max} \left\{ \frac{\delta_2+q_3}{q_2}, 1+\alpha \xi \right\}$, the equilibrium point $ \Tilde{E_1} = (0,y_1^*,0,y_2^*)$ is globally asymptotically stable.
\end{proof}

\section{Single Patch Analysis and Bifurcation Structure}
\label{bifurcation_Section}
We now restrict our attention to a single-patch version of the model to conduct a detailed bifurcation analysis. In this reduced setting, we consider the dynamics of the pest and predator populations within one patch in the presence of additional food, but in the absence of inter-patch movement.\\
The simplified system is given by
\begin{equation}   \label{model}
\left\{\begin{array}{l}
\displaystyle \frac{dx}{dt}=x \left (1-\frac{x}{k_p}\right )-\frac{x y}{1+x+\alpha \xi}=\displaystyle r(x)\,[F(x)-y],\\
\displaystyle \frac{dy}{dt}=\epsilon \left ( \frac{x+\xi}{1+x+\alpha \xi} \right)\ y - c\, \xi\, y^p=y\, [\epsilon \, h(x)-c\, \xi \, y^{(p-1)}], \ 1<p \leq 2
\end{array}\right.
\end{equation}
where 
\begin{align}
    r(x)&:=\frac{x}{1+x+\alpha\xi}, \quad 
    F(x):=\left(1-\frac{x}{k_p}\right)(1+x+\alpha\xi), \quad \text{and} \quad 
    h(x):=\frac{x+\xi}{1+x+\alpha\xi}
\end{align}
To make our system biologically admissible, we consider \eqref{model} in the region $\mathbb{R}_+^2=\{(x,y)\, |\, x\geq0,\, y\geq0\}$. Also, it can be seen that the region
\begin{equation}
    \mathcal{R}=\left\{ (x, y) \in \mathbb{R}_+^2 \;\middle|\; 0 \leq x \leq k_p,\; 0 \leq y \leq \left( \frac{\epsilon (k_p + \xi)}{(1 + k_p + \alpha \xi)\, c\, \xi} \right)^{\frac{1}{p - 1}} \right\}
\end{equation}
of system \eqref{model} is positively invariant and bounded, ensuring that solutions starting within it remain nonnegative and bounded for all $t\geq0$.
 
Three equilibrium points lie on the boundary $\partial \mathcal{R}$: $D_0 = (0, 0)$, representing the extinction of both species; $D_{k_p} = (k_p, 0)$, representing the extinction of the predator population; and $D_y = (0, y)$, where $y=\left(\frac{\epsilon}{c (\alpha  \xi +1)}\right)^{\frac{1}{p-1}}$, representing a pest-free state where the predator persists due to the presence of additional food and is well-defined as $c>0$, $\alpha>0$ and $\xi>0$. The equilibrium $D_y$ exists for $1 < p \leq 2$.
\subsection{Linear Analysis}
\begin{figure}
 \begin{subfigure}{.32\textwidth}
\includegraphics[width = 5cm, height=5cm]{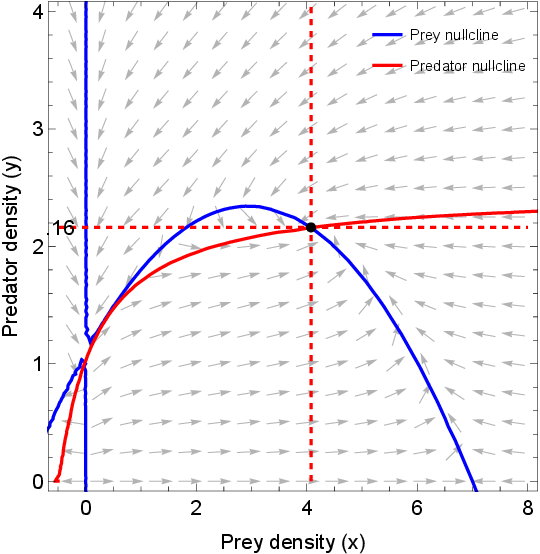}
\caption{}
\label{fig:one_int}
\end{subfigure}
  \begin{subfigure}{.32\textwidth}
  \includegraphics[width= 5cm, height=5cm]{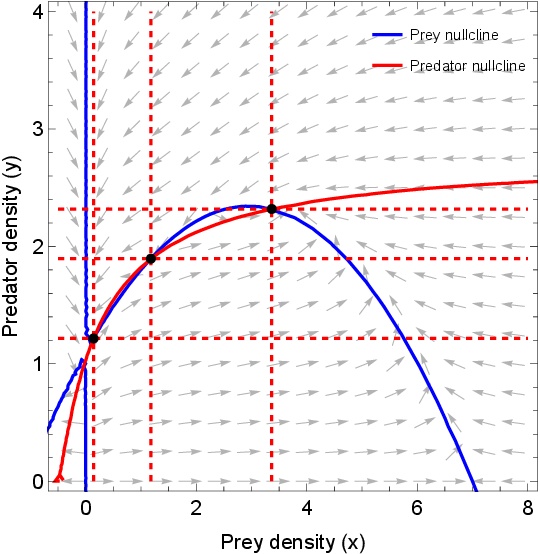}
  \caption{}
\label{fig:three_int}
  \end{subfigure}
  \begin{subfigure}{.32\textwidth}
  \includegraphics[width= 5cm, height=5cm]{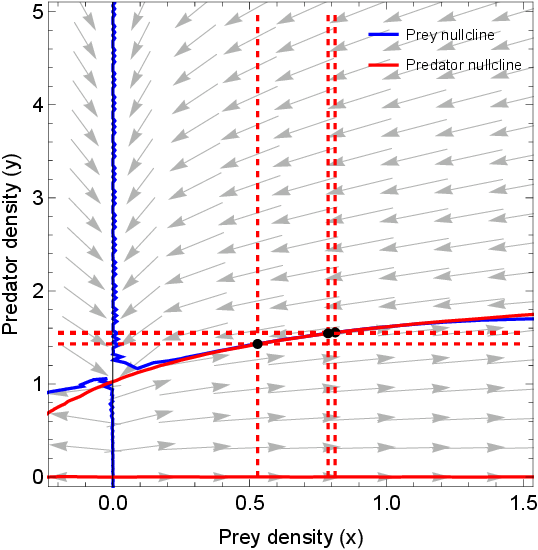}
  \caption{}
\label{fig:three_int_p2}
\end{subfigure}
\caption{\textbf{Existence of interior equilibria via nullcline intersections.} 
Parameters used: $\alpha=0.2$, $\xi=0.5$, $c=0.20456$, $\epsilon=0.23153$. 
\ref{fig:one_int} $k_p=7$, $p=1.9$: one interior equilibrium; 
\ref{fig:three_int} $k_p=7$, $p=1.8$: three interior equilibria; 
\ref{fig:three_int_p2} $k_p=4.32857$, $p=2$: three interior equilibria within a restricted range of prey density. 
Boundary equilibria are omitted.}
\label{Fig:int_equi_existance}
\end{figure}

The Jacobian matrix for \eqref{model} at any equilibrium point $(x^*,y^*)$ is given by:
\begin{equation} \label{vm}
\mathbb{M}(x^*,y^*)=
\begin{bmatrix}
r(x^*)F'(x^*)+r'(x^*)(F(x^*)-y^*) & -r(x^*) \\
\epsilon h'(x^*)y^*  &  \epsilon h(x^*)-p\, c\, \xi (y^*)^{(p-1)}
\end{bmatrix},
\end{equation}
For the interior equilibrium $E = (x^*, y^*)$ with $y^* = F(x^*)$, we have
\begin{itemize}
    \item $\operatorname{tr}\left(\mathbb{M}(x^*, F(x^*))\right)=$ $r(x^*)F'(x^*)-(p-1)\, c\, \xi\, (F(x^*))^{(p-1)}$,
    \vskip 3mm
    \item $\det\left(\mathbb{M}(x^*, F(x^*))\right)=$ $r(x^*)F(x^*)[\epsilon h'(x^*)-(p-1)c\xi (F(x^*))^{(p-2)}F'(x^*)]$.
\end{itemize}
To locate the positive interior equilibrium, we solve the equation
\begin{equation}
\epsilon\, \frac{x^* + \xi}{1 + x^* + \alpha \xi} - c\, \xi\, \left( \left(1 - \frac{x^*}{k_p} \right)(1 + x^* + \alpha \xi) \right)^{p - 1} = 0 \label{eq:g_zero}
\end{equation}
The existence of interior equilibria can be visualized via nullcline intersections ( see Fig. \ref{Fig:int_equi_existance}). Depending on the parameter regime, the system admits either a single positive equilibrium (see Fig. \ref{fig:one_int}) or up to three distinct equilibria (see Figs. \ref{fig:three_int}–\ref{fig:three_int_p2}).\\
Define
\begin{equation}
g(x^*) := \epsilon\, \frac{x^* + \xi}{1 + x^* + \alpha \xi} - c\, \xi\, \left( \left(1 - \frac{x^*}{k_p} \right)(1 + x^* + \alpha \xi) \right)^{p - 1}
\end{equation}
Then, differentiating $g(x^*)$ yields
\begin{equation}
\label{eq:gprime}
\begin{aligned}
g'(x^*) = \frac{\epsilon \left(1 + (\alpha - 1)\, \xi \right)}{(1 + x^* + \alpha\, \xi)^2} - 
\frac{c\, (p - 1)\, \xi\, (-1 - 2x^* + k_p - \alpha\, \xi)}{k_p} \left( \left(1 - \frac{x^*}{k_p} \right)(1 + x^* + \alpha \xi) \right)^{p - 2}
\end{aligned}
\end{equation}
Solving equation~\eqref{eq:g_zero} for \(c\), we obtain:
\begin{equation}
c = - \frac{\epsilon\, (x^* - k_p)\, (x^* + \xi)}{k_p\, \xi} \left( \left(1 - \frac{x^*}{k_p} \right)(1 + x^* + \alpha \xi) \right)^{-p}
\end{equation}
Substituting this expression for \(c\) into~\eqref{eq:gprime} and determinant equation, we get:
\begin{equation}\label{eq.4.9}
\begin{aligned}
    g'(x^*) &= \frac{\epsilon \left(1 + (\alpha - 1)\, \xi \right)}{(1 + x^* + \alpha\, \xi)^2} - 
    \frac{(p - 1)\, \epsilon\, \, (x^* + \xi)\, (1 + 2x^* - k_p + \alpha\, \xi)}{(x^* - k_p)  (1 + x^* + \alpha\, \xi)^2}, \\
    \det\left(\mathbb{M}\right) &= x^* \left(1 - \frac{x^*}{k_p} \right) \left[
    \frac{\epsilon \left(1 + (\alpha - 1)\, \xi \right)}{(1 + x^* + \alpha\, \xi)^2} - 
    \frac{(p - 1)\, \epsilon\, \, (x^* + \xi)\, (1 + 2x^* - k_p + \alpha\, \xi)}{(x^* - k_p)  (1 + x^* + \alpha\, \xi)^2}
    \right]
\end{aligned}
\end{equation}
Thus, the determinant of the Jacobian at the interior equilibrium can be written compactly as:
\begin{equation} \label{eq:det_compact}
\det(\mathbb{M}) = x^* \left(1 - \frac{x^*}{k_p} \right) g'(x^*)
\end{equation}

\begin{rem}
As illustrated in Fig. \ref{fig:three_int_p2}, the existence of three interior equilibria for $p=2$ has also been reported in a related predator-prey system that incorporates a predator death term (see \citep{PWB23}). In our setting, a similar multiplicity of equilibria is present, but without the predator's death term.
\end{rem}
\subsection{Bogdanov-Takens bifurcation: cusp of order 2}
For the system \eqref{model} to exhibit a Bogdanov–Takens (BT) bifurcation at an interior equilibrium $(x^*,y^*)$, the Jacobian $\mathbb{M}$  at $(x^*,y^*)$ must have a double zero eigenvalue in a single Jordan block. This implies that both the determinant and the trace of the Jacobian matrix $\mathbb{M}$ vanish simultaneously. That is, \[
\det\left(\mathbb{M}\right) = 0 \quad \text{and} \quad \operatorname{tr}\left(\mathbb{M}\right) = 0.
\]
From \eqref{eq:det_compact},
\begin{align}
    \det\left(\mathbb{M}\right)&=x^* \left(1 - \frac{x^*}{k_p} \right)g'(x^*)=0\quad \implies g'(x^*)=0
\end{align}
Solving $g'(x^*)=0$ for $k_p$, we obtain the critical value:
\begin{equation}
    k_p=\frac{
2(p-1)\, {x^*}^2 + (p-1)\, \xi\, (1 + \alpha\, \xi) + 
x^* \left(p-2 + p(2 + \alpha)\, \xi - (1 + 2\alpha)\, \xi \right)
}{ (p-1)\, x^* + p\, \xi - \alpha\, \xi-1}:=k_p^*
\end{equation}
Next, the trace of the Jacobian matrix is given by
\begin{align}
    \operatorname{tr}\left(\mathbb{M}\right)=- \frac{(p-1)\, \epsilon\, (x^* + \xi)}{1 + x^* + \alpha\, \xi} + \frac{x^* (1+( \alpha-1)\, \xi)}{2\, (p-1)\, {x^*}^2 + (p-1)\, (1 + \alpha\, \xi)\, \xi + x^*\, \left( p-2 + (-1 - 2\alpha + p(2 + \alpha))\, \xi \right)}
\end{align}
Solving $\operatorname{tr}\left(\mathbb{M}\right)=0$ for $\epsilon$ yields the critical value:
\begin{equation}
    \epsilon=\frac{
x^*\left(1 + (\alpha-1)\, \xi\right)(1 + x^* + \alpha\, \xi)}{(p-1)(x^* + \xi) \left[2(p-1)\, {x^*}^2 + (p-1)\,(1 + \alpha\, \xi)\, \xi + x^*\left(p-2 + (-1 - 2\alpha + p(2 + \alpha))\, \xi\right)\right]}:=\epsilon^*
\end{equation}
To ensure that both parameters \( k_p^* \) and \( \epsilon^* \) are positive, the following conditions must hold for \( x^* > 0 \), \( \alpha > 0 \), \( \xi > 0 \), and \( 1 < p \leq 2 \):
\begin{equation}
\label{eq:conditions_case1}
\begin{aligned}
\text{For } 0 < x^* \leq 1: \quad
&\text{If } 0 < \alpha < 1, \quad \frac{x^* - 1}{\alpha - 2} < \xi < \frac{1}{1 - \alpha}, \quad \text{and} \quad p > \frac{1 + x^* + \alpha \xi}{x^* + \xi}; \\
&\text{If } 1 \leq \alpha < 2, \quad \xi > \frac{x^* - 1}{\alpha - 2}, \quad \text{and} \quad p > \frac{1 + x^* + \alpha \xi}{x^* + \xi}.
\end{aligned}
\end{equation}

\begin{equation}
\label{eq:conditions_case2}
\begin{aligned}
\text{For } x^* > 1: \quad
&\text{If } 0 < \alpha < 1, \quad 0 < \xi < \frac{1}{1 - \alpha}, \quad \text{and} \quad p > \frac{1 + x^* + \alpha \xi}{x^* + \xi}; \\
&\text{If } 1 \leq \alpha \leq 2, \quad \xi > 0, \quad \text{and} \quad p > \frac{1 + x^* + \alpha \xi}{x^* + \xi}; \\
&\text{If } \alpha > 2, \quad 0 < \xi < \frac{x^* - 1}{\alpha - 2}, \quad \text{and} \quad p > \frac{1 + x^* + \alpha \xi}{x^* + \xi}.
\end{aligned}
\end{equation}

We now show that when $(k_p,\epsilon)=(k_p^*,\epsilon^*)$, the equilibrium $E=(x^*, F(x^*))$ is a cusp singularity of codimension-$2$. To establish this, we require the following lemma \citep{perko2013differential}.
\begin{lemma}\label{perko_result}
Any system of the form
    \begin{eqnarray}
\left\{\begin{aligned}
\dot{x}= &\ y +A_1x^2 + A_2xy + A_3y^2 + O(|(x,y)^3|),\\
\dot{y}=&\ B_1x^2+B_2xy + B_3y^2+O(|(x,y)^3|)
\end{aligned}\right.
\end{eqnarray}
is equivalent to the system
 \begin{eqnarray}
\left\{\begin{aligned}
\dot{x}= &\ y  + O(|(x,y)^3|),\\
\dot{y}=&\ B_1x^2+(B_2+2A_1)xy +O(|(x,y)^3|)
\end{aligned}\right.
\end{eqnarray}
\end{lemma}

\begin{thm}
    For any choice of \( x^* > 0 \), \( \alpha > 0 \), \( \xi > 0 \), and \( 1 < p \leq 2 \) satisfying conditions \eqref{eq:conditions_case1}-\eqref{eq:conditions_case2}, the equilibrium $E=(x^*, F(x^*))$ is a cusp singularity of codimension-$2$ precisely when $(k_p,\epsilon)=(k_p^*,\epsilon^*)$.
    \label{thm:BT2} 
\end{thm}
\begin{proof}
    We begin by shifting coordinates via the affine transformation $x_1=x-x^*$ and $y_1=y-F(x^*)$, which brings the equilibrium $E=(x^*,F(x^*))$ to the origin. Expanding the system in a Taylor series around $E$, we obtain the following reduced system:
\begin{eqnarray}\label{normal}
\left\{\begin{aligned}
\dot{x_1}= &\ r(x^*)F'(x^*)x_1 -r(x^*)y_1 +\frac{r(x^*)F''(x^*)+2r'(x^*)F'(x^*)}{2}x_1^2\\
&\ -r'(x^*)x_1y_1 + O(|(x_1,y_1)^3|),\\
\dot{y_1}=&\ \epsilon F(x^*)h'(x^*)x_1-(p-1)c\xi (F(x^*))^{(p-1)}y_1+ \frac{\epsilon F(x^*) h''(x^*)}{2}x_1^2 \\
&\  +\epsilon h'(x^*)x_1y_1 -\frac{p(p-1)c\xi (F(x^*))^{p-2}}{2}y_1^2 +O(|(x_1,y_1)^3|),
\end{aligned}\right.
\end{eqnarray}
Under the Bogdanov–Takens (BT) bifurcation conditions,
\begin{equation}\label{eq:BT_cond}
    r(x^*)F'(x^*)-(p-1)c\xi (F(x^*))^{(p-1)}=0\quad \text{and}\quad r(x^*)F(x^*)[\epsilon h'(x^*)-(p-1)c\xi (F(x^*))^{(p-2)}F'(x^*)]=0,
\end{equation} 
system \eqref{normal} simplifies to
\begin{eqnarray}\label{eq:reduced_system}
\left\{\begin{aligned}
\dot{x_1}= &\ a_0x_1 -\frac{a_0^2}{b_0}y_1 +a_1x_1^2+a_2x_1y_1 + O(|(x_1,y_1)^3|),\\
\dot{y_1}=&\ b_0x_1 -a_0y_1 +b_1x_1^2+b_2x_1y_1+b_3y_1^2 + O(|(x_1,y_1)^3|),
\end{aligned}\right.
\end{eqnarray}
where the coefficients are given by
\begin{align*}
    a_0&=r(x^*)F'(x^*),\quad a_1=\frac{r(x^*)F''(x^*)+2r'(x^*)F'(x^*)}{2}, \quad a_2=-r'(x^*),\\
    b_0&=\epsilon F(x^*)h'(x^*), \quad b_1 = \frac{\epsilon F(x^*) h''(x^*)}{2}, \quad
b_2 = \epsilon\, h'(x^*), \quad 
b_3 = -\frac{p\, r(x^*)\, F'(x^*)}{2 F(x^*)}.
\end{align*}
From the first BT condition in equation \eqref{eq:BT_cond}, we have
\begin{equation*}
    r(x^*)F'(x^*)=(p-1)c\xi (F(x^*))^{(p-1)}>0 \quad \text{for $1<p\leq2$}
\end{equation*}
Since $r(x^*)>0$, this implies $F'(x^*)>0$ and hence
\begin{equation*}
    a_0=r(x^*)F'(x^*)\neq0
\end{equation*}
Similarly, the second BT condition in equation \eqref{eq:BT_cond} yields
\begin{align*}
    \epsilon F(x^*) h'(x^*)=r(x^*)(F'(x^*))^2>0
\end{align*}
which directly implies
\begin{equation*}
    b_0=\epsilon F(x^*) h'(x^*)\neq0
\end{equation*}
Hence, to bring the system \eqref{eq:reduced_system} to a more canonical form, we apply the transformation 
\begin{equation}
    x_1=x_2+\frac{y_2}{a_0} \quad \text{and} \quad y_1=\frac{b_0}{a_0}x_2
\end{equation}
Under this change of variables, the system becomes
\begin{eqnarray}\label{eq:reduced_system2}
\left\{\begin{aligned}
\dot{x_2}= &\ y_2 +c_{20}x_2^2+c_{11}x_2y_2+c_{02}y_2^2 +  O(|(x_2,y_2)^3|),\\
\dot{y_2}=&\ d_{20}x_2^2+d_{11}x_2y_2+d_{02}y_2^2 + O(|(x_2,y_2)^3|),
\end{aligned}\right.
\end{eqnarray}
where
\begin{align*}
c_{20}&=\left(\frac{a_0^2 b_1+a_0 b_0 b_2 + b_0^2 b_3}{a_0 b_0}\right), \quad c_{11}=\left(\frac{2 a_0 b_1 + b_0 b_2}{a_0 b_0}\right), \quad c_{02}=\frac{b_1}{a_0 b_0},\\
d_{20}&=\left(a_0 a_1 + a_2 b_0 - \frac{a_0^2 b_1}{b_0} - a_0 b_2 - b_0  b_3\right), \quad d_{11}=\left(2 a_1 + \frac{a_2 b_0}{a_0} - \frac{2 a_0 b_1}{b_0}- b_2\right), \quad d_{02}=\left(\frac{a_1 b_0 - a_0 b_1}{a_0 b_0}\right)
\end{align*}
Using lemma \eqref{perko_result}, the system \eqref{eq:reduced_system2} is transformed to the standard normal-form:
\begin{eqnarray}\label{codim2}
\left\{\begin{aligned}
\dot{x_2}= &\ y_2 + O(|(x_2,y_2)^3|),\\
\dot{y_2}=&\ \beta_1 x_2^2
+ \beta_2 x_2y_2 + O(|(x_2,y_2)^3|)
\end{aligned}\right.
\end{eqnarray}
where the normal-form coefficients are given by
\begin{align}
    \beta_1&=\left(a_0  a_1 + a_2  b_0 - \frac{a_0^2 b_1}{b_0} - a_0  b_2 - b_0  b_3\right)\notag\\ 
    &=\frac{(p-2)\, r(x^*)^2F'(x^*)^3 }{2F(x^*)}+\frac{ r(x^*)^2F'(x^*)F''(x^*) }{2}-\frac{\epsilon^* r(x^*)F(x^*)h''(x^*) }{2},\\
    \beta_2&=\frac{2a_0a_1+b_0a_2+a_0b_2+2b_0b_3}{a_0}\notag\\
    &=r'(x^*)F'(x^*)-(p-1)\, \epsilon^*\, h'(x^*)+ r(x^*)F''(x^*).
\end{align}
Therefore, if $\beta_1\beta_2\neq0$, the $E=(x^*,F(x^*))$ is a cusp singularity of codimension-$2$, and the system exhibits a codimension-$2$ Bogdanov–Takens bifurcation.
\end{proof}
\begin{rem}
    The signs of the normal-form coefficients $\beta_1$ and $\beta_2$ depend on the underlying parameter values $(x^*, \alpha, \xi, p)$ and hence on the expressions for $k_p^*$ and $\epsilon^*$. Thus, both $\beta_1$ and $\beta_2$ can be either positive or negative depending on the specific choice of these parameters.
\end{rem}

\begin{figure}
   \begin{subfigure}{.5\textwidth}
  \includegraphics[width= 8cm, height=4.3cm]{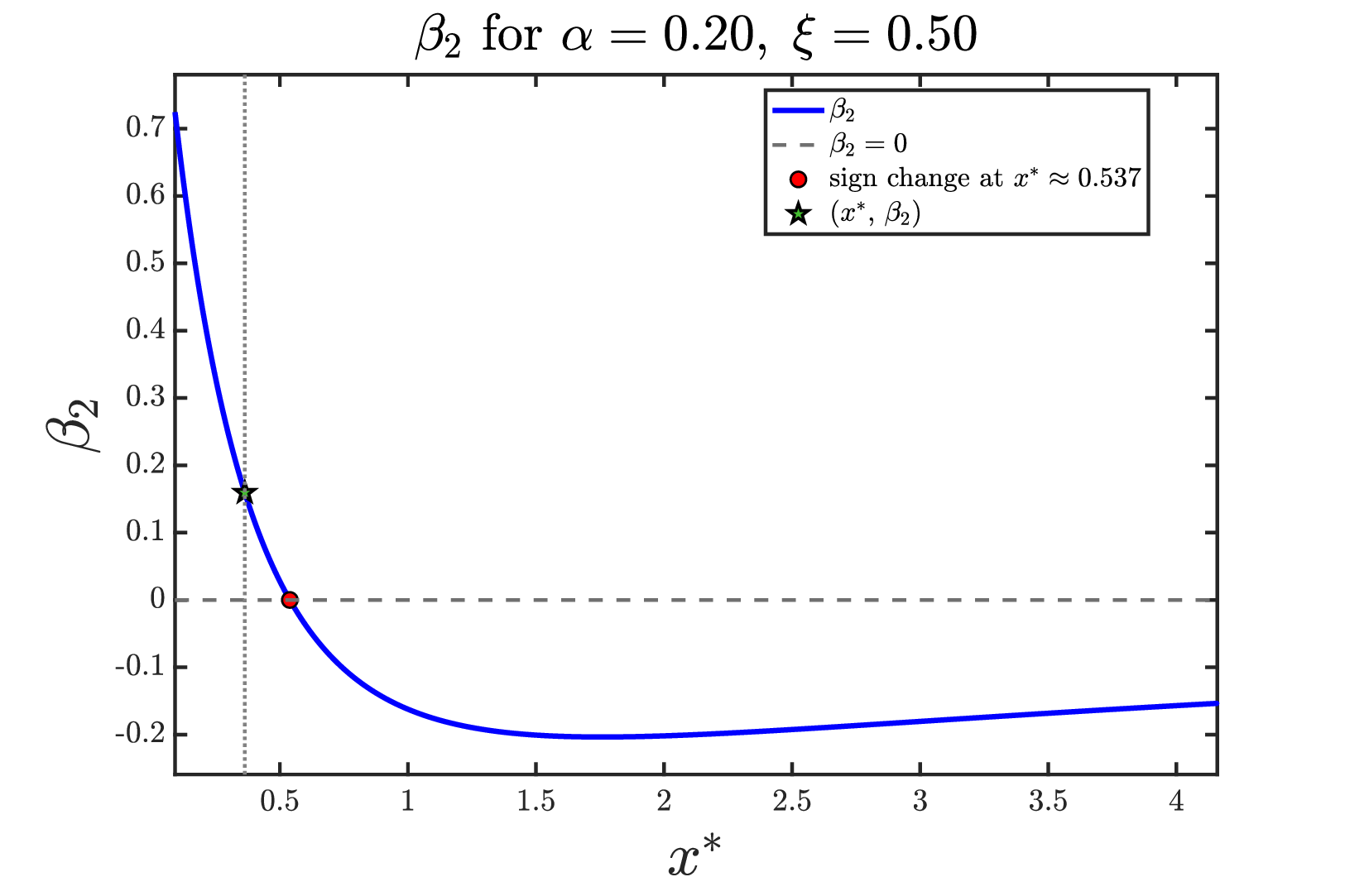}
  \caption{}
 \label{fig:beta2_plot}
  \end{subfigure}
   \begin{subfigure}{.5\textwidth}
  \includegraphics[width= 8cm, height=4.4cm]{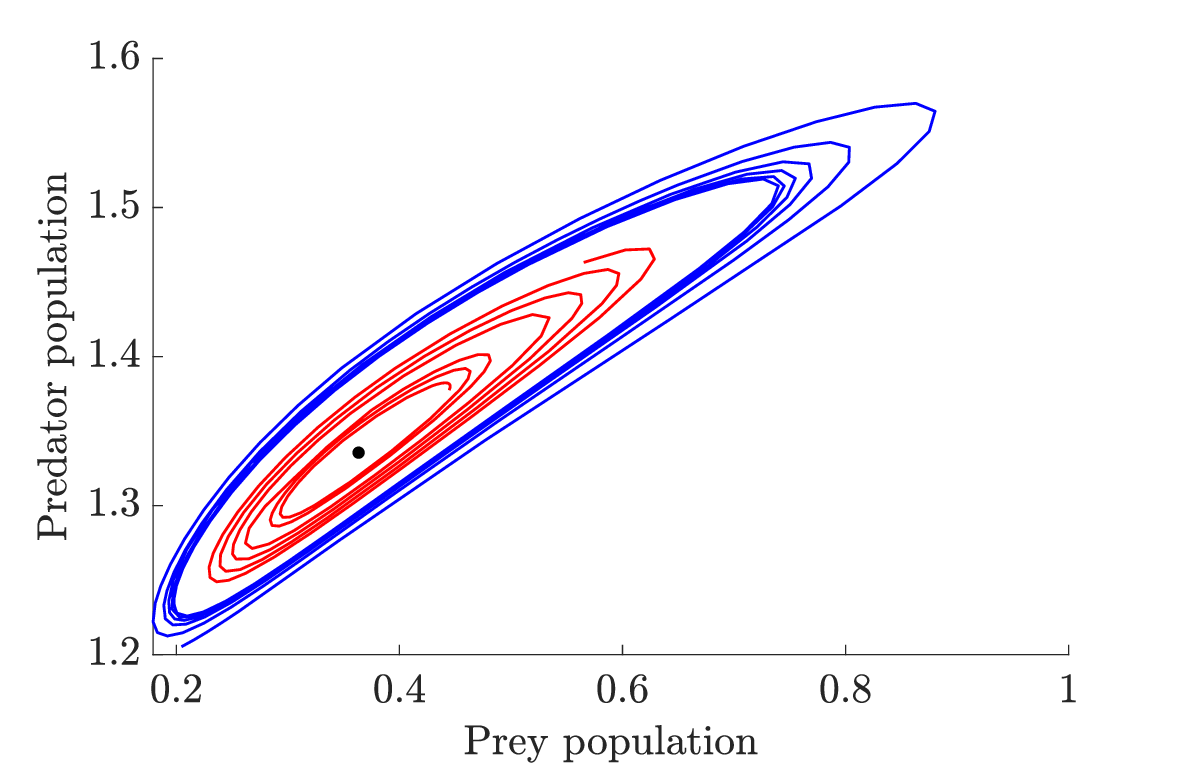}
  \caption{}
\label{fig:stable_limit_cycle}
 \end{subfigure}
\caption{Fig. \ref{fig:beta2_plot} 
shows the behavior of the normal-form coefficient $\beta_2$ for $\alpha=0.20$ and $\xi=0.50$. The dashed line marks $\beta_2=0$; the red dot indicates the sign change at $x^*\approx0.537$. The green star highlights the equilibrium abscissa used in Fig. \ref{fig:stable_limit_cycle}, $x^*=0.36336$, where $\beta_2=0.159438>0$ (vertical dotted line). Fig. \ref{fig:stable_limit_cycle} Phase portrait of system \eqref{model} with $(k_p^*,\epsilon^*,c,p)=(4.16,0.23153,0.20456,2)$ and the same $(\alpha,\xi)$ as in Fig. \ref{fig:beta2_plot}. Trajectories from the I.C.s, $(0.20465,1.20529)$ (blue) and $(0.44465,1.37748)$ (red), converge to a unique attracting periodic orbit (stable limit cycle), while the interior equilibrium $E=(0.36336,1.33550)$ (black dot) is an unstable focus enclosed by the cycle. Together, the figures indicate that for these parameters the model lies in the regime $\beta_{2}>0$ and exhibits a stable limit cycle surrounding the unstable equilibrium.}
\label{Fig:BT-bifurcation}
\end{figure}

\section{Discussion \& Conclusion}
\label{Discussion and Conclusion}
In this study, we develop and analyze a two-patch biological control model - this is relevant to model current innovative agricultural practice in a landscape containing a prairie strip ($\Omega_1$) and a crop field ($\Omega_2$). This is motivated by recent innovative developments in landscape management, such as the STRIPS program. We posit the following assumptions: (i) only predators can move between patches; while prey stay in the crop field; (ii) predator movement between $\Omega_1$ and $\Omega_2$ is modeled via drift \eqref{eq:model_drift_cann} and dispersal \eqref{eq:patch_model2}; (iii) AF initiates generalized competition among predators within the prairie strip. The effect of drift and dispersal between the two patches has been studied, and the biological implications of the results have also been analyzed.

\par 
 When the predators move via drift, away from $\Omega_1$ and into $\Omega_2$, then the complete pest extinction state in both patches is globally stable under certain parametric conditions via Theorem \ref{thm:E1_global_unidirectional}; (see Fig. \ref{nullcline_complete_extinction}).  The conditions outlined in Theorem \ref{thm:E1_global_unidirectional} indicate that the intraspecific competition rate $c$ needs to be less than a certain threshold, suggesting that the competition among predators must be weak enough and that drift into the crop field must be fast enough to compensate for predators' mortality for the result to hold. The pest extinction only in the crop field is also globally stable via Theorem \ref{thm:E2_global_unidirectional};   (see 
Fig. \ref{x2_extinction_cann_nullcline_drift}) suggests that the predator competition should be greater than a threshold so that the prey-predator can coexist in $\Omega_1$. Also, the drift into $\Omega_2$ should be sufficient to eradicate the pest from the crop field.  Similar outcomes were observed when predators dispersed between the patches in both directions. Complete pest eradication is globally asymptotically stable via Theorem \ref{e1_patch1_dispersal}; (see Fig. \ref{com_ext_disp_nullcline}) indicating that the predator density in $\Omega_1$ should be high enough to keep a constant predation pressure on the pest. The interaction of predators between patches also drives predators in $\Omega_2$ above a certain level, which helps eliminate pests from the crop field.  Lemma \ref{lem:E2_existence_dispersal} provides the existence of pest extinction only in the crop field state and the
related local stability via Lemma \ref{lem:E2_stability_dispersal} for model \ref{eq:patch_model2}; (see Fig. \ref{x2_extinction_cann_nullcline_dispersal}).  
\par 
In section \ref{bifurcation_Section}, we conducted a detailed bifurcation analysis for the single-patch model \ref{model} in the presence of additional food supplements to reveal how predator-prey dynamics change with biological parameters. In particular, we established the conditions for the existence of a codimension-$2$ Bogdanov-Takens (BT) bifurcation via Theorem \ref{thm:BT2}. This BT bifurcation corresponds to the parameter regime where the Jacobian at an interior equilibrium point has a double zero eigenvalue, signaling the onset of rich and intricate local behavior. It also acts as an organizing center from which multiple dynamical phenomena can emerge. In its unfolding, one typically observes the appearance of saddle-node bifurcations and homoclinic loops — each contributing to qualitative changes such as the birth or destruction of limit cycles, bistability, and sharp transitions. Fig. \ref{Fig:BT-bifurcation} shows the emergence of a stable limit cycle, which corresponds to long-term, sustained oscillations in pest and predator populations, representing ecological scenarios where neither species goes extinct nor settles to a steady state. The presence of limit cycles in the vicinity of the BT point is essential for developing some robust pest control strategies that are resilient to disturbances in the environment. In this work, we restricted our analysis to the generic case $\beta_1\beta_2\neq0$, which leads to a codimension-2 Bogdanov-Takens bifurcation. The degenerate case $\beta_1\beta_2=0$ may involve higher codimension bifurcations and more complex local dynamics, but lies outside the scope of the current work and remains a direction for future study.   Furthermore, in the current work, we consider the regime $1< p \leq 2$ for the $-y^p$ generalized competition-type term. It remains as future work to investigate the regime $0 < p < 1$. This is a more delicate regime to handle, due to the possibility of finite time extinction dynamics, and its ensuing, often counterintuitive effect on the competitive dynamics \cite{PAT21, banerjee2025two}.

\par 
The summary of our findings illustrates that intraspecific competition among predators can be beneficial for pest extinction in a crop field, which is the primary area of concern from a biological control perspective. This result is of importance to the control of invasive pests such as the soybean aphid, which has plagued soybean crops since its first detection in the United States in Wisconsin in July 2000 \cite{ragsdale2004soybean}. The main predator of the soybean aphid is the Asian Beetle (\emph{Harmonia Axyridis}), which is known to have evolved several aggressive traits, making it a fierce inter and intraspecific competitor \cite{osawa2011ecology}. Thus, a possible nuanced strategy is to plant certain AF that \emph{Harmonia Axyridis} prefers in prairie strips adjoining soybean crop fields. Herein, as increased precipitation is predicted in the Midwestern US \cite{lee23}, drift between adjoining fields, due to vectors such as flooding, becomes increasingly important. Also, a possible direction of current and future interest is to consider interspecific competition among various biological control agents, such as predators, parasitoids, pathogens, and combinations thereof \cite{verma2025towards}. These could include the effects of intraguild predation. As noted earlier, we have proved the existence of a BT2 bifurcation in the AF (single) patch setting. 
We expect this result to hold in the case of linear drift. In addition, the presence of limit cycles that occur in the vicinity of the BT point is relevant to maintaining a cyclical population of predators. In terms of application to the soybean aphid and its chief predator \emph{Harmonia Axyridis}, one can consider the $\epsilon-k_{p}$ parameters. Next, we can infer $\epsilon$, the conversion efficiency of the predator from the literature - then choose an appropriate $k_{p}$, which essentially governs the ``size" of the prairie strip, so that we stay in a feasible BT2 regime. This method would enable an AF patch of appropriate size, so that we could maintain a cyclical population of predators, dispersing/drifting into an adjoining soybean field, to target the aphid. All in all, we believe our results have value in designing tactics and strategies relevant to the practical control of various current invasive pests. These strategies are particularly relevant considering the increasing effects of climate change, and the evolution of dispersal strategies for pests, as habitat fragmentation continues to increase the patchiness of agricultural landscapes.
\section{Funding}
UV, KG, and RP acknowledge valuable partial support by the Agricultural and Food Research Initiative grant no. 2023-67013-39157 from the USDA National Institute of Food and Agriculture.
\section{Appendix}
\label{Appendix}
\subsection{Proof of Theorem \ref{pos_inv_drift}}
\begin{proof}
From the differential equations of $x_1,x_2$ we have, $\dot{x_1}\rvert_{x_1=0}=0, \dot{x_2} \rvert_{x_2=0}=0$. And same reasoning applies to $\dot{y_1}$ we have $\dot{y_1}\rvert_{y_1=0} = 0$. Now for  $\dot{y_2}$ we have $\dot{y_2}= q_2 y_1$ if $y_2=0$ and this implies $\dot{y_2}\rvert_{y_2=0} \geq 0$ if $y_1 \geq 0$. Therefore, by Theorem $A.4$ in \citep{thieme2018mathematics}, the model \eqref{eq:patch_model2} is positive invariant in $\mathbb{R}_+^4$. 
    \label{proof:pos_inv_drift}
\end{proof}

\subsection{Proof of Lemma \ref{lem:E1_existence_unidirectional}}
\begin{proof}
\label{proof_E1_existence_unidirectional}
From the nullcline $ \Dot{y_2}=0$ we have,
\begin{equation}
    - \delta_2 y_2^* + q_2 y_1^*   = 0 \ \implies 
y_2^* = \dfrac{q_2 y_1^* }{\delta_2}
\label{y2_pest_ext_comp}
\end{equation}
From the nullcline $\Dot{y_1}=0$ we have, 
\begin{equation*}
y_1^*\ \left \{  \frac{\epsilon_1  \xi}{1+\alpha \xi}  - q_1 -c \xi (y_1^*)^{p - 1} \right \} = 0
\end{equation*}
\begin{equation}
   \frac{  \epsilon_1 \xi}{1+\alpha \xi} - q_1 =  c \xi (y_1^*)^{p - 1}
\ \implies \ 
   y_1^* = \left( \dfrac{\epsilon_1 \xi-  q_1(1+\alpha \xi) }{c\xi(1+\alpha \xi)}\right)^{\frac{1}{p-1}}, \ p \neq 1
   \label{explicit_y1_comp_ext}
\end{equation}
Thus, the equilibrium point $\hat{E_1} = (0,y_1^*,0,y_2^*)$ exists if  $\xi > \frac{ q_1 }{\epsilon_1- \alpha q_1 }$ and,  {$ \epsilon_1 > \alpha q_1 $}.
\end{proof}

\subsection{Proof of Lemma \ref{lem:stability_pest_ext_drift}}
\begin{proof}
\label{proof:stability_pest_ext_drift}
Using \eqref{explicit_y1_comp_ext}, the general Jacobian matrix  \eqref{general_jacobian_uni} at $(0,y_1^*,0,y_2^*)$ becomes, 
\begin{equation*} 
\hat{J_1} = \begin{bmatrix}
1- \frac{y_1^*} {(1 +\alpha \xi)}   & 0  & 0 & 0 \vspace{0.25cm}
  \\ 
\frac{\epsilon_1  (1 + (\alpha - 1) \xi) \ y_1^*}{(1+\alpha \xi)^2} & c \xi (1-p)(y_1^*)^{p-1} & 0 & 0 \vspace{0.25cm}
\\
0 & 0 & 1  - y_2^* & 0 \vspace{0.25cm}
  \\
0 & q_2 & \epsilon_2 y_2^* &  - \delta_2 \\
\end{bmatrix}
\end{equation*} 
Now the characteristic equation is given as,
\begin{equation*}
 (c \xi (1-p)(y_1^*)^{p-1}-\lambda) \left( 1- y_2^* - \lambda \right) (-\delta_2-\lambda) \left(1- \dfrac{y_1^*} {1 +\alpha \xi} - \lambda \right)  = 0
 \end{equation*}
\begin{equation*}
\lambda_1 = c \xi (1-p)(y_1^*)^{p-1} \ < 0 \Leftrightarrow  p>1 \ \text{(which is true)}, \quad \lambda_2 =  1  - y_2^* < 0 \Leftrightarrow  y_2^*>1 \Leftrightarrow y_1^* > \frac{\delta_2}{q_2}, 
\end{equation*}
\begin{equation*}
\lambda_3  = - \delta_2 \ (< 0), \quad
\lambda_4 = 1  -  \dfrac{y_1^* } {1 +\alpha \xi} < 0 \Leftrightarrow y_1^
*>(1+\alpha \xi) 
\end{equation*}
Therefore,  $\hat{E_1} = (0,y_1^*,0,y_2^*)$ is locally asymptotically stable if $y_1^*>\text{max} \left\{ \frac{\delta_2}{q_2}, 1+\alpha \xi \right\}$.
\end{proof}

\subsection{Proof of Lemma \ref{lem:E2_existence_unidirectional}}
\begin{proof}
\label{proof_E2_existence_unidirectional}
From the nullcline  $ \Dot{y_2}=0$ we have,
 \begin{equation} 
     y_2^* = \frac{q_2 y_1^*}{\delta_2 } 
     \label{x2_0_eq_1_uni}
 \end{equation}
From the nullcline $ \Dot{x_1}=0$,
\begin{equation*}
    x_1^*\ \left \{ \left (1-\frac{x_1^*}{k_p} \right)-\frac{ y_1^*}{1+x_1^*+\alpha \xi} \right \} = 0
\end{equation*}
either $ x_1^* = 0$ or,
\begin{equation}
  y_1^* =  \left(1-\frac{x_1^*}{k_p}\right) (1+x_1^*+\alpha \xi) >0 \ \text{when,} \ x_1^* < k_p 
    \label{x2_0_eq_2_uni}
\end{equation}
Now, using the predator nullcline in patch $\Omega_1$,

\begin{equation*}
   y_1^*\ \left \{ \epsilon_1 \left( \frac{x_1+\xi}{1+x_1+\alpha \xi} \right )  - q_1  - c \xi  (y_1^*)^{p-1} \right \} = 0
\end{equation*}
$\therefore$ either $y_1^* = 0$ or,

\begin{equation}
\label{x2_0_eq_3_uni}
   \epsilon_1 \left ( \frac{x_1^* +\xi}{1+x_1^* +\alpha \xi} \right)  - q_1 - c \xi  (y_1^*)^{p-1} = 0
\end{equation}
Solving the above equation for $x_1^*$, we have
\begin{equation*}
     x_1^* = \frac{A \ \left (1+\alpha \xi \right ) - \epsilon_1 \xi }{\epsilon_1 - A} 
\end{equation*}
where $A$ is defined by,
\begin{equation*}
   A= q_1 + c \xi  (y_1^*)^{p-1} >0
\end{equation*}
Thus, the equilibrium point $\hat{E_2} = (x_1^*,y_1^*,0,y_2^*)$ exists if  $ \epsilon_1 \xi < A (1+\alpha \xi)$ and {$ \epsilon_1 >  A $}.
\end{proof}

\subsection{Proof of Lemma \ref{lem:E2_stability_unidirectional}}
\begin{proof}
\label{proof_E2_stability_unidirectional}
The general Jacobian matrix  \eqref{general_jacobian_uni} at $(x_1^*,y_1^*,0,y_2^*)$ becomes, 

\begin{equation*} 
\hat{J_2} = \begin{bmatrix}
\frac{ x_1^* y_1^*} {(1+x_1^* +\alpha \xi)^2}  - \frac{ x_1^*}{k_p} & \frac{- x_1^*}{1+x_1^* +\alpha \xi} & 0 & 0 \vspace{0.25cm}
  \\ 
\frac{\epsilon_1  (1 + (\alpha - 1) \xi) \ y_1^*}{(1+x_1^* +\alpha \xi)^2} &  c\xi (1-p)(y_1^*)^{p-1} & 0 & 0 \vspace{0.25cm}
\\
0 & 0 & 1  - y_2^* & 0 \vspace{0.25cm}
  \\
0 & q_2 & \epsilon_2 y_2^* &  - \delta_2 \\
\end{bmatrix}
\end{equation*} 
\vspace{0.25cm}
The characteristic polynomial is, 
\begin{equation*}
\left(1- y_2^* - \lambda \right) \left[\left( \dfrac{ x_1^* y_1^*} {(1+x_1^* +\alpha \xi)^2} - \dfrac{ x_1^*}{k_p}  - \lambda\right)\left(\lambda^2 + \lambda (\delta_2 - c \xi (1-p)(y_1^*)^{p-1}) - c \xi \delta_2(1-p)(y_1^*)^{p-1}  \right) \right] 
\end{equation*}
\begin{equation*}
-\left(1- y_2^* - \lambda \right) \left[\dfrac{ x_1^*\left(\delta_2 + \lambda\right)}{1+x_1^* +\alpha \xi} \left(\dfrac{\epsilon_1  (1 + (\alpha - 1) \xi) \ y_1^*}{(1+x_1^* +\alpha \xi)^2}\right) \right] 
\end{equation*}
\normalsize
\begin{equation*}
\text{Let, }  A_2 = \dfrac{ x_1^* y_1^*} {(1+x_1^* +\alpha \xi)^2}  - \dfrac{ x_1^*}{k_p} , \quad  B_2 = \dfrac{ - x_1^*}{1+x_1^* +\alpha \xi}, \quad  C_2 = \dfrac{\epsilon_1  (1 + (\alpha - 1) \xi) \ y_1^*}{(1+x_1^* +\alpha \xi)^2},
\end{equation*}

\begin{equation*}
E_2=  c \xi (1-p)(y_1^*)^{p-1}, \quad D_2 = 1-y_2^* \text{  where } B_2  < 0
\end{equation*}
Now the characteristic equation becomes,
\begin{equation*}
\left(D_2- \lambda \right)   \{ \lambda^3 + \lambda^2(\delta_2 - A_2-E_2)+ \lambda(A_2 E_2-A_2\delta_2-B_2 C_2-E_2 \delta_2) - B_2 C_2 \delta_2 +E_2 A_2 \delta_2 \}= 0
\end{equation*}
To satisfy the Routh–Hurwitz stability criteria for all negative roots, we should have the following conditions:
\begin{equation}
\begin{aligned}
\label{R-H for x2_0_unidirectional}
& 1-y_2^*>0,\quad  \delta_2>A_2 +E_2,\ \quad   A_2(E_2-\delta_2) > +B_2C_2 + E_2 \delta_2,\ E_2 A_2- B_2C_2>0 \ \&\\
& (\delta_2-A_2-E_2)(A_2E_2-A_2\delta_2-B_2C_2-E_2 \delta_2)>\delta_2(B_2C_2-E_2 A_2) 
\end{aligned}
\end{equation}
Thus, if all the conditions mentioned in equation \eqref{R-H for x2_0_unidirectional} are satisfied, then the lemma is proved.
\end{proof}
\subsection{Proof of Theorem \ref{positively invariant}}
\begin{proof}
From the differential equations of $x_1,x_2$ we have, $\dot{x_1}\rvert_{x_1=0}=0, \dot{x_2} \rvert_{x_2=0}=0$. And looking at $\dot{y_1}$ we have $\dot{y_1}= q_4 y_2$ if $y_1=0$ and this implies $\dot{y_1}\rvert_{y_1=0} \geq 0$ if $y_2 \geq 0$, applying the same logic for $\dot{y_2}$ we have, $\dot{y_2}\rvert_{y_2=0} \geq 0$ if $y_1 \geq 0$. Therefore, by Theorem $A.4$ in \citep{thieme2018mathematics}, the model \eqref{eq:patch_model2} is positive invariant in $\mathbb{R}_+^4$. Let $D$ be the region where, $D=\{(x_1,y_1,x_2,y_2) \in \mathbb{R}_+^4\, |\, x_1\geq0,\, y_1\geq0, \ x_2\geq0, y_2\geq0\}$.

\label{proof:positve_dispersal}
\end{proof}

\subsection{Proof of Lemma \ref{lem:E1_existence_dispersal}}
\begin{proof}
From the nullcline $ \Dot{y_2}=0$ we have, 

\begin{equation}
     y_2^* = \dfrac{q_2}{\delta_2 + q_3} \ y_1^*
     \label{x1_x2_0_eq1}
 \end{equation}
and from $\Dot{y_1} = 0 $,
\begin{equation*}
 y_1^* \left\{ \frac{\epsilon_1 
 \xi}{1+\alpha \xi} \  - q_1  - c \xi  (y_1^*)^{p-1} \right \} + q_4 y_2^* = 0 ,
\end{equation*}
Using the value of $y_2^*$ from \eqref{x1_x2_0_eq1},
\begin{equation*}
 y_1^* \left\{ \frac{\epsilon_1 
 \xi}{1+\alpha \xi} \  - q_1  - c \xi  (y_1^*)^{p-1} +\dfrac{q_4 q_2}{\delta_2 + q_3}  \right \}  = 0 ,
\end{equation*}
either $y_1^* = 0 $ or,
\begin{equation*}
\left\{ \frac{\epsilon_1 
 \xi}{1+\alpha \xi} \  - q_1  - c \xi (y_1^*)^{p-1} +\dfrac{q_4 q_2}{\delta_2 + q_3}  \right \}  = 0 
 \label{x1_x2_0_value} 
\end{equation*}

\begin{equation}
y_1^* = \left\{ \dfrac{1}{c \xi }\left(\frac{\epsilon_1 
 \xi}{1+\alpha \xi} - q_1 +\dfrac{q_4 q_2}{\delta_2 + q_3} \right)\right \}^{\frac{1}{p-1}}, \ p \neq 1
\label{x1_x2_0_eq2}    
\end{equation}
So for positivity of $y_1^*$  we need, $q_1 < \frac{\epsilon_1 
 \xi}{1+\alpha \xi}  +\frac{q_4 q_2}{\delta_2 + q_3}$. Solving this for $\xi$ gives, $\xi > \frac{q_1 - \frac{q_4 q_2}{\delta_2 + q_3}}{\epsilon_1 - \alpha \left(q_1 - \frac{q_4 q_2}{\delta_2 + q_3}\right)},$ 
 $ \text{provided } \epsilon_1 - \alpha \left(q_1 - \frac{q_4 q_2}{\delta_2 + q_3}\right) > 0 $. Thus the equilibrium 
$ \Tilde{E_1} = (0,y_1^*,0,y_2^*)$ exists if $\xi >  \frac{B}{\epsilon_1-\alpha B}, \ \text{and}, \ \epsilon_1-\alpha B>0 $ where $B=q_1 - \frac{q_4 q_2}{\delta_2 + q_3}$.
\label{proof:E1_existence_dispersal}
\end{proof}

\subsection{Proof of Lemma \ref{lem:stability_pest_ext_dispersal}}
\begin{proof}
Using equation \eqref{x1_x2_0_eq2}, the general Jacobian matrix \eqref{general_jacobian_dispersal} at $(0,y_1^*,0,y_2^*)$ becomes, 
\begin{equation} 
\tilde{J_1} = \begin{bmatrix}
1  -  \frac{y_1^* } {\left(1+\alpha \xi\right)}  & 0 & 0 & 0 \vspace{0.25cm}
  \\ 
\frac{\epsilon_1  \left(1 + \left(\alpha - 1\right) \xi\right) \ y_1^*}{(1+\alpha \xi)^2} &   \ \ c\xi 
(1-p)(y_1^*)^{p-1} - \frac{q_4 q_2}{\delta_2 + q_3} & 0 & q_4 
\\
0 & 0 & 1 - y_2^* & 0 
\vspace{0.25cm}
  \\
  0 & q_2 & \epsilon_2 y_2 &  - \delta_2 - q_3
  \end{bmatrix}
\label{jacobian_comp_ext_dispersal}
 \end{equation}
 Now the characteristic equation is given as,
 \begin{equation*}
 \left(c \xi (1-p)(y_1^*)^{p-1}- \dfrac{q_4 q_2}{\delta_2 + q_3}-\lambda\right) \left( 1- y_2^* - \lambda \right) (-\delta_2-q_3-\lambda) \left(1- \dfrac{y_1^*} {1 +\alpha \xi} - \lambda \right)  = 0
 \end{equation*}
 \begin{equation*}
\lambda_1 = c \xi (1-p)(y_1^*)^{p-1} -\dfrac{q_4 q_2}{\delta_2 + q_3}\ < 0, \text{since} \ 1-p<0 \ \text{then,} \ y_1^* > \left( \frac{q_4 q_2}{c \, \xi \, (1 - p) (\delta_2 + q_3)} \right)^{\frac{1}{p - 1}}, \text{holds by existence}
\end{equation*}
\begin{equation*}
 \ \lambda_2 =  1  - y_2^* < 0 \Leftrightarrow  y_2^*>1 \Leftrightarrow y_1^* > \frac{\delta_2+q_3}{q_2}, \quad \lambda_3  = - \delta_2-q_3 \ (< 0), \quad
\lambda_4 = 1  -  \dfrac{y_1^* } {1 +\alpha \xi} < 0 \Leftrightarrow y_1^
*>(1+\alpha \xi) 
\end{equation*}
\label{proof:stability_pest_ext_dispersal}
Therefore,  $\Tilde{E_1} = (0,y_1^*,0,y_2^*)$ is locally asymptotically stable if  $y_1^*>\text{max} \left\{ \frac{\delta_2+q_3}{q_2}, 1+\alpha \xi \right\}$.
\end{proof}

\subsection{Proof of Lemma \ref{lem:E2_existence_dispersal}}
\begin{proof}
From the nullcline  $ \Dot{y_2}=0$ we have,

\begin{equation}
     y_2^* = \dfrac{q_2}{\delta_2 + q_3} \ y_1^*
     \label{x2_0_eq1}
 \end{equation}
Using the nullcline $ \Dot{x_1}=0$,
     
     \begin{equation*}
    x_1^*\ \left \{ 1-\frac{x_1^*}{k_p} -\frac{ y_1^*}{1+x_1^*+\alpha \xi} \right \} = 0
\end{equation*}
either $ x_1^* = 0$ or,
\begin{equation}
  y_1^* =  \left(1-\frac{x_1^*}{k_p}\right) (1+x_1^*+\alpha \xi) 
  \label{x2_0_eq_2}
\end{equation}
So, for  $y_1^* >0 \ \text{we should have,} \ x_1^* < k_p$.
Now, using the nullcline  $ \Dot{y_1}=0$, and substituting the value of $y_2^*$ from \eqref{x2_0_eq1} we have,
\begin{equation*}
   y_1^*\ \left \{ \epsilon_1 \left( \frac{x_1^*+\xi}{1+x_1^*+\alpha \xi} \right ) - c \xi  (y_1^*)^{p-1}- q_1   + \dfrac{q_4 q_2} {\delta_2 + q_3}  \right \} = 0
\end{equation*}
$\therefore$ either $y_1^* = 0$ or,
\begin{equation}
   \epsilon_1 \left ( \frac{x_1^* +\xi}{1+x_1^* +\alpha \xi} \right)  - c \xi  (y_1^*)^{p-1}- q_1   + \dfrac{q_4 q_2} {\delta_2 + q_3}  = 0
    \label{x2_0_eq_3}
\end{equation}
Solving for $x_1^*$ we have, 
\begin{equation}
     \epsilon_1 \left ( \frac{x_1^* +\xi}{1+x_1^* +\alpha \xi} \right) =  c \xi  (y_1^*)^{p-1}+ q_1   -\dfrac{q_4 q_2} {\delta_2 + q_3}
\ \implies \ 
     x_1^* = \frac{D \ \left (1+\alpha \xi \right ) - \epsilon_1 \xi }{\epsilon_1 - D} 
      \label{x2_0_eq_4}
 \end{equation}
where $D$ is defined by,
\begin{equation*}
 D= c \xi  (y_1^*)^{p-1}+ q_1   -\dfrac{q_4 q_2} {\delta_2 + q_3}\  
  \end{equation*}
Thus, the equilibrium point $\Tilde{E_2} = (x_1^*,y_1^*,0,y_2^*)$ exists if  $\epsilon_1 \xi < D(1+\alpha \xi)$ and {$ \epsilon_1 >  D $}.
\label{proof_E2_existence_dispersal}
\end{proof}

\subsection{Proof of Lemma \ref{lem:E2_stability_dispersal}}
\begin{proof}
Using \eqref{x2_0_eq_3} and \eqref{x2_0_eq_2} the general Jacobian matrix \eqref{general_jacobian_dispersal} at $(x_1^*,y_1^*,0,y_2^*)$ becomes,
\begin{equation} 
\tilde{J_2} = \begin{bmatrix}
\frac{ x_1^* y_1^*} {(1+x_1^* +\alpha \xi)^2}  - \frac{ x_1^*}{k_p} & \frac{- x_1^*}{1+x_1^* +\alpha \xi} & 0 & 0 \vspace{0.25cm}
  \\ 
\frac{\epsilon_1  \left(1 + \left(\alpha - 1\right) \xi\right) \ y_1^*}{(1+x_1^* +\alpha \xi)^2} &    c\xi (1-p)(y_1^*)^{p-1} - \frac{q_4 q_2} {\delta_2 + q_3}& 0 & q_4 
\\
0 & 0 & 1 - y_2^* & 0
 \vspace{0.25cm}
  \\
  0 & q_2 & \epsilon_2 y_2 &  - \delta_2 - q_3
  \end{bmatrix}
\label{x2_0_jacobian_dispersal}
 \end{equation}
\begin{equation*}
\text{Let, }     \tilde{D_2}= 1-y_2^*, \ \tilde{A_2} = \dfrac{ x_1^* y_1^*} {(1+x_1^* +\alpha \xi)^2}  - \dfrac{ x_1^*}{k_p} , \ \tilde{B_2} = \dfrac{ - x_1^*}{1+x_1^* +\alpha \xi}, \ \tilde{C_2} = \dfrac{\epsilon_1  (1 + (\alpha - 1) \xi) \ y_1^*}{(1+x_1^* +\alpha \xi)^2},
\end{equation*}
\begin{equation*}
    \tilde{E_2} =  c\xi (1-p)(y_1^*)^{p-1} - \dfrac{q_4 q_2} {\delta_2 + q_3}\ \text{  where }\  \tilde{B_2}  < 0. \ \text{The Jacobian matrix can now be written as,} \ 
\end{equation*}
\begin{equation} 
\tilde{J_2} = \begin{bmatrix}
\tilde{A_2} & \tilde{B_2} & 0 & 0 \vspace{0.25cm}
  \\ 
\tilde{C_2} &   \tilde{E_2}& 0 & q_4 
\\
0 & 0 & \tilde{D_2} & 0
 \vspace{0.25cm}
  \\
  0 & q_2 & \epsilon_2 y_2 &  - \delta_2 - q_3
  \end{bmatrix}
\label{x2_0_jacobian_dispersal_simplified}
 \end{equation}
The characteristic polynomial comes out as:
\begin{equation*}
 (\tilde{D_2}-\lambda )\ \left[ \left(\tilde{A_2}- \lambda \right)   \left\{(\tilde{E_2}-\lambda)(-\delta_2 - q_3-\lambda)-q_4 q_2 \right \}+ \tilde{B_2}\tilde{C_2}   \left(\delta_2 + q_3+ \lambda \right)  \right]
\end{equation*}
Further simplification gives us the following characteristic equation, 
\begin{equation*}
    (\tilde{D_2}-\lambda )\ \left[ \left(\tilde{A_2}- \lambda \right)   \left\{\lambda^2+ \lambda(\delta_2 + q_3-\tilde{E_2})-c\xi (1-p)(y_1^*)^{p-1}(\delta_2+q_3) \right \}+ \tilde{B_2}\tilde{C_2}   \left(\delta_2 + q_3+ \lambda \right)  \right]=0
\end{equation*}
Let $\tilde{F_2}= \delta_2 + q_3$ and, $\tilde{G_2}=c\xi (1-p)(y_1^*)^{p-1}(\delta_2+q_3) $. Rewriting the above equation gives, 
\begin{equation*}
\setlength{\jot}{10pt}
\begin{aligned}
   & (\tilde{D_2}-\lambda )\ \left[ \left(\tilde{A_2}- \lambda \right)   \left\{\lambda^2+ \lambda(\tilde{F_2}-\tilde{E_2})- \tilde{G_2} \right \}+ \tilde{B_2}\tilde{C_2}   \left(\tilde{F_2}+ \lambda \right)  \right]=0\\
   &\implies \lambda^4 - \lambda^3\left( \tilde{D_2}+\tilde{E_2}-\tilde{F_2}+\tilde{A_2}\right) +  \lambda^2 \left(\tilde{D_2}(\tilde{E_2}-\tilde{F_2}+\tilde{A_2}) + \tilde{A_2}(\tilde{E_2}-\tilde{F_2})- \tilde{G_2}-\tilde{B_2}\tilde{C_2}\right) \\
   & + \lambda \left(\tilde{A_2}\tilde{D_2}  (\tilde{F_2}-\tilde{E_2})+ \tilde{D_2}(\tilde{G_2}+\tilde{B_2}\tilde{C_2})-\tilde{B_2}\tilde{C_2}\tilde{F_2} 
   + \tilde{A_2}\tilde{G_2}\right) + \tilde{B_2}\tilde{C_2}\tilde{D_2}\tilde{F_2} - \tilde{A_2}\tilde{D_2}\tilde{G_2} = 0 \nonumber  \\
   &\implies \lambda^4 +A_3 \lambda^3 + A_2 \lambda^2 + A_1 \lambda + A_0  = 0  \\
   &\text{where, }A_3 = - \left( \tilde{D_2}+\tilde{E_2}-\tilde{F_2}+\tilde{A_2}\right) , \quad A_2 = \left(\tilde{D_2}(\tilde{E_2}-\tilde{F_2}+\tilde{A_2}) + \tilde{A_2}(\tilde{E_2}-\tilde{F_2})- \tilde{G_2}-\tilde{B_2}\tilde{C_2}\right),\\
   & \  A_1 = \left(\tilde{A_2}\tilde{D_2}  (\tilde{F_2}-\tilde{E_2})+ \tilde{D_2}(\tilde{G_2}+\tilde{B_2}\tilde{C_2})-\tilde{B_2}\tilde{C_2}\tilde{F_2} + \tilde{A_2}\tilde{G_2}\right) ,  \quad A_0 = \tilde{B_2}\tilde{C_2}\tilde{D_2}\tilde{F_2} - \tilde{A_2}\tilde{D_2}\tilde{G_2} \\
&\text{To satisfy Routh–Hurwitz stability criteria for all negative roots, the following conditions should hold,}\\
& A_3 > 0,  A_2 > 0 , A_1 > 0, A_0 > 0, \text{ and }  A_3A_2A_1 > A_1^2 + A_3^2A_0.\nonumber
    \end{aligned}
\end{equation*}

\begin{equation} 
\text{for }  A_0 > 0 , \tilde{D_2}(\tilde{B_2}\tilde{C_2}\tilde{F_2} - \tilde{A_2}\tilde{G_2}) > 0 \implies \tilde{B_2}\tilde{C_2}\tilde{D_2}\tilde{F_2} > \tilde{A_2}\tilde{D_2}\tilde{G_2}
    \label{R-H for x2_0}
\end{equation}
\begin{equation}
\text{for }  A_1 > 0, \left(\tilde{A_2}\tilde{D_2}  (\tilde{F_2}-\tilde{E_2})+ \tilde{D_2}(\tilde{G_2}+\tilde{B_2}\tilde{C_2})-\tilde{B_2}\tilde{C_2}\tilde{F_2} + \tilde{A_2}\tilde{G_2}\right)>0
    \label{R-H for x2_1}
\end{equation}
\begin{equation}
\text{for }  A_2 >0, \left(\tilde{D_2}(\tilde{E_2}-\tilde{F_2}+\tilde{A_2}) + \tilde{A_2}(\tilde{E_2}-\tilde{F_2})- \tilde{G_2}-\tilde{B_2}\tilde{C_2}\right)>0
    \label{R-H for x2_2}
\end{equation}
\begin{equation}
    \text{for }  A_3 > 0,  - \left( \tilde{D_2}+\tilde{E_2}-\tilde{F_2}+\tilde{A_2}\right)  >0 \implies  \left( \tilde{D_2}+\tilde{E_2}-\tilde{F_2}+\tilde{A_2}\right)  <0
        \label{R-H for x2_3}
\end{equation}
\begin{equation}
    \text{and then }  A_3A_2A_1 > A_1^2 + A_3^2A_0
    \label{R-H for x2_0_final}
\end{equation}
Thus, if all conditions mentioned from \eqref{R-H for x2_0} - \eqref{R-H for x2_0_final} hold the equilibrium point   $\Tilde{E_2} = (x_1^*,y_1^*,0,y_2^*)$ is locally asymptotically stable. 
\label{proof_E2_stability_dispersal}
\end{proof}

\bibliographystyle{elsarticle-num} 
\bibliography{references}
\end{document}